\documentstyle[epsf,11pt]{article}
\begin{document}
\title       {Solving Gauge Field Theory \\
                 by Discretized Light-Cone Quantization } 
\author{Hans-Christian Pauli \\ 
               Max-Planck-Institut f\"ur Kernphysik \\
               D-69029 Heidelberg \\ }
\date{24 January 1996; revised 31 July 1996}
\maketitle
\begin{abstract}
The canonical front form Hamiltonian for non-Abelian  SU(N)
gauge theory in 3+1 dimensions is mapped on an effective 
Hamiltonian which acts 
only in the Fock space of one quark and one antiquark.
The approach is non-perturbative and exact. It is based on
Discretized Light-Cone Quantization and the Method of
Iterated  Resolvents. The method resums the diagrams of
perturbation theory to all orders in the coupling constant
and is free of Tamm-Dancoff truncations in the Fock-space.
Emphasis is put on  dealing accurately with the many-body
aspects of gauge field theory. Pending future renormalization 
group analysis  the running coupling is derived to all orders 
in the bare coupling constant.~--- 
The derived effective interaction has an amazingly simple 
structure and is gauge invariant and frame independent.
It is solvable on a small computer like a  work station. 
The many-body amplitudes can be retrieved self-consistently 
from these solutions, by quadratures without solving another 
eigenvalue problem. 
The structures found allow also for developing simple
phenomenological models consistent with non-Abelian 
gauge field theory.
\end{abstract}
\vfill 
\noindent  \ Preprint MPIH-V25-1996 
\newpage \tableofcontents \newpage 
\section {Introduction and Motivation} 
\label {sec:1} 

One of the most important tasks in hadron physics is to
calculate and  understand the mass spectrum and the wave
functions of physical hadrons from a covariant quantum field
theory.  The wave functions encode all the properties which are 
needed for a phenomenological description of experiments. 
Wave functions and probability amplitudes have a natural 
appearance in a Hamiltonian approach.

But the Hamiltonian bound state problem is 
notoriously difficult in a field theory. Procedures like those of
Schwinger and  Dyson or  of Bethe and Salpeter are not easy 
to cope with in practice as reviewed recently in 
\cite{lsg91}.  
Actually, the difficulties were considered so enormous that
the Hamiltonian approach was given up in the Fifties alltogether 
in favor of Feynman's action oriented approach. 
Modern  successors like the Lattice Gauge Theories
\cite{mac93,wei94} govern the scene, 
maturing  from infancy in recent times \cite{nrq95}.  
Phenomenological  models
\cite{goi85,dkm91,neu93,eiq94}  are closer to experiment
and have the different objective to classify the bulk of 
empirical data. They leave little doubt that a heavy 
meson contains primarily a pair of  
constituent quarks and {\em not} an infinity of sea particles 
as suggested by the perturbative treatment of quantum 
field theory. 

How can one reconcile these models 
particularly the constituent quark model (CQM) with the
quantum field theory of chromodynamics (QCD)?
There are several reasons why the front form~\cite{dir49}
of Hamiltonian dynamics \cite{leb80} is one of the very few
candidates,  see \cite{leb81,brl89,brp91} or \cite{gla95}. 
Particularly
the simple vacuum and the simple boost properties
\cite{brp91,pab85a,pab85b,epb87} confront with the
`complicated vacuum' and the `dynamical boosts' 
of the conventional approach, the instant form.
These aspects are stressed also in the arguments of
Wilson~\cite{wil89} and collaborators \cite{wwh94}.
Nowadays we know that even the front form vacuum
is not simple~\cite{hks92,vap92,piv94}, but  still 
{\em simplier} than in the instant form:  The problem
can at least be formulated~\cite{kap93a,pkp95}.

The success of discretized light-done quantization (DLCQ) 
particularly in 1+1 dimensions  
has stirred hope that its apparent simplicities
carry over to the physical 3+1 dimensions 
\cite{tab91,kap92,kpw92,wol95}. But this meets problems, 
among them: The Hamiltonian matrix increases 
exponentially fast with the particle number of the
Fock states and with the number of transversal momenta.
Truncating the Fock space to 2 particles like
Tamm \cite{tam45}   and Dancoff \cite{dan50} invokes
perturbation theory, violates gauge invariance and generates
non-integrable singularities. One must resort to {\it ad hoc}
procedures to make things working \cite{kpw92}. 
Truncating to 3 particles \cite{kap92,wol95}, 
the numerical results are inconclusive
as to test an onset of convergence. 

Even worse than that: Thus far, it is obscure how 
{\em any Hamiltonian} including the one on the light cone
could be subjected to a renormalization group analysis. 
This and the possibly large coupling
constant has motivated Wilson \cite{wwh94}
to give up the one-to-one connection with canonical field
theory and to propose a radically new procedure in which
the renormalization properties of front form operators play 
the crucial role. The problem is a very deep one and appears
also in Quantum Electrodynamics (QED). 
In QED, the smallness of the
coupling constant obscures the fact that the simple 
Coulomb potential between two point charges has 
not been derived from field theory thus far with other than
perturbative methods. Although higher order effects are 
expected  to be small, and are indeed so, see for example 
\cite{boy78,byg85,sty84} and \cite{cal78},  
no manifestly non-perturbative and closed analytical 
procedures are at hand. This is remarkable. In QCD the
problems are only accentuated due to the larger coupling 
constant. 

By practical considerations one must reduce the matrix
dimension of the Hamiltonian. 
But the reduction to comparatively small matrices 
is  more than merely a technical issue. It becomes 
{\em a matter of principle}.  When analyzing the source of the
difficulty one realizes that the many-body aspects of a 
field theory have not been mastered thus far: Precisely those
are treated in perturbation theory and precisely those are
responsible for the large matrix dimensions. 
It is here where the present work supposedly contributes. 
The many-body aspects are kept in the form as they want to
appear, as resolvents, and these {\em are not expanded} 
as perturbative series. 

Our point is then:
One should  solve first the many-body aspects of the 
canonical Hamiltonian, which has not been done thus far,
and then think on the problems of renormalization. 
This is not in conflict with \cite{wwh94}, but we emphasize
a different aspects of the problem.
One is faced then with problems similar to conventional
many-body physics, problems  as one meets them in the
theory of atoms, nuclei or of  solids.
Remarkably, one can carry out this programme
essentially without assumption, and still arrive at  a
solvable equation. As a matter of fact it is also simple.

At the core of this work is a new method, 
the `method of iterated resolvents'  
to be introduced in section~\ref{sec:4}.
It allows to do `perturbation theory to all orders' 
and `perturbation theory in medium' 
without a smallness parameter. 
For to be specific, some of the ingredients 
of earlier work \cite{brp91},  particularly the 
Lagrangian  and the DLCQ-Fock-space for QCD, 
are summarized shortly in section~\ref{sec:2}.
Section~\ref{sec:3} summarizes the theory of effective
interactions as known from the literature \cite{mof50} and  
displays why the Tamm-Dancoff approach 
\cite{tam45,dan50} is bound to fail.
The results of section~\ref{sec:4} are then applied to QCD
in section~\ref{sec:6}.
At the end of many formal
manipulations one takes the continuum limit. 
The effective  interaction of a quark and an
antiquark  will then turn out as a sum of two terms which
have an intuitively appealing  interpretation:
(1) The effective potential $U$, generated by the exchange 
of one effective gluon which is  absorbed either by the same
or by the other quark; and (2) The effective annihilation
interaction $U_a$, where the quark and antiquark annihilate 
into two effective gluons. The potential $U$ is derived 
explicitly in section~\ref{sec:7}, and section~\ref{sec:8}
summarizes the results including  a broader discussion 
on their possible future application.
\section {The front form Hamiltonian for Quantum Chromodynamics}
\label {sec:2}

Both  in non-relativistic quantum mechanics and in field theory, 
the Hamiltonian operator propagates the system in time.
In a covariant theory the concept of `time' can be 
generalized, since the space-time parametrization is arbitrary.
But following Dirac \cite{{dir49}}, 
there are no more than three standard forms 
how to choose generalized time and the corresponding
Hamiltonian dynamics: the `instant', the `front' and
the `point' form. In this section we shall summarize  
\cite{leb80,leb81,brl89,brp91} 
the front form Hamiltonian for  QCD, actually for SU(N), 

The time-like coordinate is chosen as $ x^+ = t+z$, 
the space-like coordinates as 
$\vec x = (x, y, t-z) \equiv (\vec x _{\!\bot} , x^-)$.
The four-vector of space-time is thus 
$x^\mu = (x^+, \vec x _{\!\bot} , x^-)$.
In QCD the vector potentials 
$ ({\bf A}  ^\mu)_{cc^\prime}$ are $3\times 3$ matrices, 
and the Dirac spinors carry a color index $c$, 
{\it i.e.}  $\Psi _{\alpha,c}$ with $c=1,2,3$.
The Lagrangian density is given by the hermitian operator
\begin {equation} 
   \displaystyle {\cal L}  = 
 - {1\over2} {\rm Tr}  {{\bf F}^{\mu\nu} {\bf F}_{\mu \nu} }
 + {1\over2} \Bigl[ 
   \overline\Psi \bigl( i\gamma^\mu {\bf D} _\mu 
 - m_{\rm F}\bigr) \Psi  
 + {\rm h.c.} \Bigr] 
\ , \end {equation}  
expressed in terms of the covariant  derivative 
${\bf D} _\mu \equiv  \partial_\mu + ig {\bf A}_\mu$
and the color-electromagnetic fields 
${\bf F}^{\mu\nu}\equiv\partial^\mu {\bf A}^\nu - \partial^\nu {\bf A}^\mu
 + i g  \left[ {\bf A} ^\mu , {\bf A} ^\nu \right] $.  
The four components of the energy-momentum vector 
\begin {equation} 
   P ^\nu  = {1\over2} \int _\Omega \! dx^-\, d^2\vec x _{\!\bot} \ \Bigl(
   2 {\rm Tr}  {{\bf F}^{+\kappa} {\bf F}_\kappa ^{\phantom{\kappa}\nu}}
  + {1\over2} \Bigl[\overline\Psi  i\gamma^+ {\bf D} ^\nu \Psi  
  + {\rm h.c.} \Bigr]   - g ^{+\nu} {\cal L}  \ \Bigr) 
\ , \label {eq:208} \end {equation} 
{\it i.e.} $ P ^\nu = (P  ^+, \vec P  _{\!\bot}, P  ^-)$, 
are strict constants of the motion. Its space-like components 
$\vec P  _{\!\bot} =( P  ^1, P  ^2)$ and $P  ^+$
do not depend on the interaction and in momentum representation
are diagonal operators.
The time-like component $P ^-= 2 P_+ $ depends on the interaction 
and propagates the system in the light-cone time $ x^+$, {\it i.e.}  
$ i { \partial \over \partial x ^+ } \vert \Psi \rangle = 
  P _+ \vert \Psi \rangle $, 
and  therefore is the proper front form Hamiltonian  \cite{brp91}.  
The contraction of these four operators   
\begin {equation} 
      P ^\mu P _\mu  = P ^+ P ^- - \vec P  _{\!\bot} ^{\,2} 
      \qquad  \equiv H _{\rm LC} \equiv H
\label {eq:221}\end {equation} 
is Lorentz {\em invariant} and  refered to
somewhat improperly but conveniently as the 
`light-cone Hamiltonian $ H _{\rm LC} $' 
\cite{brp91}, or shortly $ H $.
One seeks a representation in which $H$ is diagonal, 
\begin {equation} 
      H \vert \Psi _i \rangle = E _i \vert \Psi _i \rangle
\ . \label {eq:222}\end {equation} 
Note that the eigenvalues $ E _i $  and matrix elements of
$H$  somewhat unusually carry the dimension of an
invariant-mass-squared.

Periodic boundary conditions on ${\cal L} $ can be realized by 
periodic boundary conditions on the vector potentials ${\bf A} ^\mu$ 
and anti-periodic boundary conditions on the spinor fields 
$\Psi _\alpha$ because  ${\cal L} $ is bilinear in the latter. 
One expands these fields into plane wave states and satifies the 
boundary conditions by  {\it discretizing} the momenta, 
hence Discretized Light-Cone Quantization (DLCQ), {\it i.e.}
\begin {eqnarray} 
    p _- 
 &=& \cases{ {\pi\over \scriptstyle L} n  , 
             &with $n  = {\scriptstyle1\over\scriptstyle2},
              {\scriptstyle3\over\scriptstyle2},\dots,
              \infty\ $ for fermion fields, \cr
              {\pi\over L} n  , 
             &with $n  = 1,2,\dots,\infty\ $ for boson 
              fields, \cr} 
\nonumber \\ 
   \quad{\rm and}\quad  \vec p _{\!\bot} 
 &=& {\pi\over L_{\!\bot}} \vec n _{\!\bot},
  \ \quad{\rm with} \ n_x,n_y = 0,\pm1,\pm2,\dots,\pm\infty
  \ \quad{\rm for}\ {\rm both}  
\ .\label {eq:210} \end {eqnarray} 
This is done at the expense of introducing two length parameters, 
$ L$ and $ L_{\!\bot}$,  which define a normalization 
volume $ \Omega \equiv 2L(2L_{\!\bot})^2$. 
More explicitly, the free fields are expanded as Fourier sums
 \begin {eqnarray} 
  \widetilde \Psi _\alpha (x)   = { 1\over \sqrt{\Omega } } 
  \sum_q { 1\over \sqrt{ p^+} }
  \left(   b_q u_\alpha (p,\lambda) e^{-ipx}  
         + d^\dagger _q v_\alpha (p,\lambda) e^{ipx} \right) , 
\nonumber \\ 
  \quad{\rm and}\quad
  \widetilde A  _\mu (x) = { 1\over \sqrt{\Omega } } 
  \sum_q { 1\over \sqrt{ p^+} }
  \left(   a _q \epsilon_\mu       (p,\lambda) e^{-ipx}  
         + a^\dagger _q \epsilon _\mu^\star (p,\lambda) e^{ipx} 
  \right) 
\ ,\label {eq:212}\end {eqnarray} 
particularly for the two transversal vector potentials
$ \widetilde A  ^i \equiv   \widetilde A  ^i _{\!\bot}$, ($i=1,2$).
Each particle is on its mass-shell $ p^\mu p_\mu = m^2$. 
Its four-momentum 
is $ p^\mu = (p^+, \vec p _{\!\bot}, p^- )$ with 
$p^- = (m^2 + \vec p ^{\,2}_{\!\bot}) / p^+ $.
Each particle state ``$q$''  is then characterized by six 
quantum numbers: 
\begin {equation} 
         q = (n, n_x, n_y, \lambda,c, f ) 
         = (p^+, \vec p_{\!\bot}, \lambda,c, f ) 
         = (x, \vec p_{\!\bot}, \lambda,c, f ) 
\ . \label {eq:214}  \end{equation} 
The first three, $(n, n_x, n_y) $, specify the space-like 
momentum and $\lambda $ the helicity  $\uparrow$ or 
$\downarrow$. 
A quark is specified further by color $c$ and flavor $f$. 
Gluons carry no flavour, and the color index $c$ is 
substituted  by the glue index $a$. 
The creation and destruction operators like $a^\dagger _q$ 
and $a _q$ create and destroy single particle states $q$, 
respectively,  and obey the usual (anti-) commutation 
relations like 
\begin {equation} 
 \big[  a _q, a^\dagger _{q^\prime} \big]  = 
   \big\{ b _q, b^\dagger _{q^\prime} \big\} = 
   \big\{ d _q, d^\dagger _{q^\prime} \big\} = 
  \ \delta _{q,q^\prime} 
\ . \end {equation}
As an advantage of DLCQ, all quantum numbers are
discrete.  One deals thus only with simple (and dimensionless) 
Kronecker symbols.
The spinors $u_\alpha $ and $v_\alpha$  
and the transversal polarization vectors $\vec \epsilon _{\!\bot} $
are the usual ones \cite{leb80} and defined in  \cite{brp91}.

A thorough treatment should include the zero modes of 
the gauge fields particularly those of $A ^+$. 
Their importance had been demonstrated \cite{kap93a,pkp95}  
for the vacuum sector. Here one deals
with the particle sectors and their massive excitations. 
Explicit calculations with \cite{pab96} and without them
\cite{vab96} yield however the same results in the continuum
limit.  The global and the gauge zero  modes
\cite{kap93a,pkp95}  are therefore discarded in the sequel, 
and the usual light-cone gauge \cite{leb80,brp91}
$         {\bf A} ^+ = 0 $ is used.
The light-cone Gauss equation, 
{\it i.e.} $\partial _\mu F^{\mu.+}_a = g J^+_a$,
see \cite{brp91} and  \cite{pkp95}, 
and the expansions in Eq.(\ref{eq:212}) 
complete the specification of all vector potentials $ {\bf A} ^\mu$.  
The space-like integrations in Eq.(\ref{eq:208}) can be carried 
out explicitly, leading essentially to Kronecker deltas.
One ends up with the light-cone energy-momenta \cite{brp91}
as operators acting in Fock space, {\it i.e.} 
$ P  ^\nu = P  ^\nu \big( a _q, a^\dagger _q, 
        b _q, b^\dagger _q, d _q, d^\dagger _q \big) $.
The various terms in the Hamiltonian are conveniently classified 
by a sum of four terms  \cite{brp91}, {\it i.e.}
\begin {equation}
      H = T + V + F + S   
\ . \label {eq:232} \end {equation} 
The kinetic energy  $T$ survives the limit of the coupling constant 
$ g $ going to zero. Since it is diagonal in Fock-space representation, 
its eigenvalue is the {\em free invariant mass squared}
of the particular Fock state. 
The {\em vertex interaction} $ V $ is the relativistic interaction
{\it per se}. It is linear in $g$ and changes the particle number by 
1 and {\em only by} 1. Matrix elements of $V$ which change the 
particle number by 3 (as in the instant form) are strictly zero 
in DLCQ:  The vacuum {\em does not fluctuate}.
The instantaneous interactions $ F $ and $ S $ are gauge 
artefacts, are consequences of working in the light-cone
gauge  and proportional to $g^2$. 
The seagull interaction $S$  conserves the particle number. 
The fork interaction $F$ changes the particle number {\em only}  by 2. 
 As illustrated below in Figure~\ref{fig:holy-1},  each block in
the  Hamiltonian matrix is therefore either the zero matrix,
or has {\em only} seagull-, or   {\em only}  vertex-, 
or  {\em only} fork-interactions, with very simple
matrix elements tabulated in \cite{brp91}. 

There is {\em one and only one} reference state which is 
annihilated by all destruction operators, 
namely the {\it Fock-space vacuum} $ \vert vac \rangle $. 
Therefore, the  Hilbert space for the single-particle creation
and annihilation   operators is the Fock space. 
It is the complete set of all possible states 
\begin {equation} 
     \vert \Phi _i \rangle = N _i
     \ b ^\dagger _{q_1} b ^\dagger _{q_2} \dots b ^\dagger _{q_N}
     \ d ^\dagger _{q_1} d ^\dagger _{q_2} \dots 
     d ^\dagger _{q_{\bar N}}
     \ a ^\dagger _{q_1} a ^\dagger _{q_2} \dots 
     a ^\dagger _{q_{\widetilde N}} \vert vac \rangle 
\  , \end {equation}
subject to  be eigenfunctions 
of  the space-like momenta, with {\em fixed eigenvalues}
$P ^+ $ and $\vec P  _{\!\bot}$, {\it i.e.}
\begin {equation}
 P ^+ = \sum _{\nu } p^+ _{\nu } = {2\pi \over L } K
\ , \quad {\rm and} \quad
 \vec P  _{\!\bot} = \sum _{\nu } (\vec p _{\!\bot} )_{\nu } 
\ . \end {equation}
The sums run over all particles  in a Fock state. 
As consequence of discretization, the Fock states are 
denumerable and orthonormal:
$\langle \Phi _i \vert \Phi _j \rangle = \delta _{ij} $.
As usual, the {\em momentum fraction} carried by the particle
is denoted by $ x  = p^+ /P ^+ $, and the sum of all fractions is
constrained to $\sum _{\nu } x _{\nu } = 1$.
Note that the Fock states can be made color neutral.
Since $P  ^+$ has only positive eigenvalues and  
since each particle has a lowest possible value of $p^+$,
the number of particles in a Fock state is limited for any fixed value 
of the {\it harmonic resolution} $ K $ \cite{pab85a,pab85b}.  
Next to the simple vacuum, this is another pecularity of DLCQ.

\def\d{$\bullet$}  \def\v{ V }  \def\b{ $\cdot$ } \def\s{ S }   \def\f{ F }
\begin{figure} [t]
\begin {center}
\caption{\label{fig:holy-1} \sl
    The Hamiltonian matrix for a SU(N)-meson. 
    The matrix elements are represented by the letters
    $S$,   $V$, and $F$, corresponding to seagull, vertex, and 
   fork-interactions, respectively. For better orientation, the 
   diagonal  blocs are marked by  ($\bullet$) and the zero 
   matrices by ($\cdot$). (In the preprint this table is replaced
   by a figure with the diagrams.)
}\vspace{1em}
\begin {tabular}  {||cc||cc||c|ccc|cccc|ccccc||}
\hline \hline 
     &   &        & N$_p$ & 
     2 & 2 & 3 & 4 & 3 & 4 & 5 & 6 & 4 & 5 & 6 & 7 & 8 
\\ 
   K & N$_p$ & Sector & n & 
     1 & 2 & 3 & 4 & 5 & 6 & 7 & 8 & 9 &10 &11 &12 &13 
\\ \hline \hline   
    1 &  2 & $ q\bar q\, $ &  1 & 
    \d &\s &\v &\f &\b &\f &\b &\b &\b &\b &\b &\b &\b 
\\ \hline
    2 &  2 & $ g\ g\ $ &  2 & 
    \s &\d &\v &\b &\v &\f &\b &\b &\f &\b &\b &\b &\b 
\\
    2 &  3 & $ q\bar q\, \ g $ &  3 & 
    \v &\v &\d &\v &\s &\v &\f &\b &\b &\f &\b &\b &\b 
\\
    2 &  4 & $ q\bar q\, q\bar q\, $ &  4 & 
    \f &\b &\v &\d &\b &\s &\v &\f &\b &\b &\f &\b &\b 
\\ \hline
    3 &  3 & $ g \ g \ g $ &  5 & 
    \b &\v &\s &\b &\d &\v &\b &\b &\v &\f &\b &\b &\b 
\\
    3 &  4 & $ q\bar q\, \ g \ g $ &  6 & 
    \f &\f &\v &\s &\v &\d &\v &\b &\s &\v &\f &\b &\b 
\\
    3 &  5 & $ q\bar q\, q\bar q\, \ g $ &  7 & 
    \b &\b &\f &\v &\b &\v &\d &\v &\b &\s &\v &\f &\b 
\\
    3 &  6 & $ q\bar q\, q\bar q\, q\bar q\, $ &  8 & 
    \b &\b &\b &\f &\b &\b &\v &\d &\b &\b &\s &\v &\f 
\\ \hline
    4 &  4 & $ g \ g \ g \ g $ &  9 & 
    \b &\f &\b &\b &\v &\s &\b &\b &\d &\v &\b &\b &\b 
\\
    4 &  5 & $ q\bar q\, \ g \ g \ g $ & 10 & 
    \b &\b &\f &\b &\f &\v &\s &\b &\v &\d &\v &\b &\b 
\\
    4 &  6 & $ q\bar q\, q\bar q\, \ g \ g $ & 11 & 
    \b &\b &\b &\f &\b &\f &\v &\s &\b &\v &\d &\v &\b 
\\
    4 &  7 & $ q\bar q\, q\bar q\, q\bar q\, \ g $ & 12 & 
    \b &\b &\b &\b &\b &\b &\f &\v &\b &\b &\v &\d &\v 
\\
    4 &  8 & $ q\bar q\, q\bar q\, q\bar q\, q\bar q\, $ & 13 & 
    \b &\b &\b &\b &\b &\b &\b &\f &\b &\b &\b &\v &\d 
\\ \hline\hline
\end {tabular}
\end {center}
\end{figure}

For to enumerate all possible Fock states for a meson with a 
fixed harmonic resolution, one needs a classification scheme.  
A possible one is displayed in Figure~\ref{fig:holy-1}. 
With Fock states being color-singlets,  the lowest possible value 
one can have for $K$ is $K=1$. 
This allows for  one $q\bar q$-pair, at the most. 
For  $K=2$, the Fock space  contains in addition the $g\,g$- and 
the $q\bar q\,g$-sectors.
For $K=4$, one has at most 8 particles, namely 4 $q\bar
q$-pairs.  In the figure all  13 Fock-space  sectors possible
for $K$ up to $4$ are denumerated by $n=1,\dots,13$.
Note that the classification of these sectors does not change 
when  $K$ is increased: one just adds more complicated
Fock space sectors to  the figure. 
The number of sectors grows  quadratically with $K$ and 
has the value $N_K =  (K+1)(K+2)/2- 2 $. 

In analogy to the figure, one can rewrite Eq.(\ref{eq:222})  
as a {\em block matrix equation}:
\begin {equation} 
      \sum _{m=1} ^{N_K} 
      \ \langle n \vert H \vert m \rangle 
      \ \langle m \vert \Psi _i \rangle 
      = E _i  \ \langle n \vert \Psi _i \rangle 
\, \qquad {\rm for\ all\ } n = 1,2,\dots,N_K 
\ .\label {eq:314} \end {equation} 
The numbers  $n$ ($m$) denumerate the sectors. 
The problem is solved if one can find the sector wave functions 
$ \langle n \vert \Psi _i \rangle $
for one or several eigenstates $ \Psi_i $. 
Note that each sector contains many individual Fock states with
different values of $x$, $\vec p  _{\!\bot}$ and $\lambda$.

Actually, right from the beginning one could have chosen the
conventional procedure. One could have written
Eq.(\ref{eq:314}) in the {\em continuum limit},
replacing sums by integrals according to 
\begin {equation} 
        \sum _{p^+={\pi\over2L}} ^{P^+}\sum _{\vec p_{\!\perp}} 
        \longrightarrow
        {P^+\over 2}\;{\Omega\over (2\pi)^3}\; \int _0^1 \!\! dx
        \int \!\! d^2\vec p  _{\!\bot}
\ . \label {eq:316} \end {equation} 
But apart from introducing innumerable integration
variables and separate summation indices like color, 
flavor and spin, one tends to loose oversight in a set of 
coupled integral equations as implied by Eq.(\ref{eq:314}). 
Since there is nothing wrong about thinking like Dirac 
in terms of matrices, we continue to do so 
because we find it more convenient.

Each sector  with a given parton number 
has arbitrarily many Fock states whose transversal momenta add up
to a given value of $\vec P  _{\!\bot} $. One needs a cut-off. 
Following  Lepage and Brodsky \cite{leb80,leb81,brl89}
the Fock space is regulated by the condition that the free 
invariant mass shall not exceed a limit, {\it i.e.} 
\begin {equation} 
      \sum _{\nu } \bigg( 
      {m ^2 + \vec p ^{\,2}_{\!\bot} \over x } \bigg) _{\nu } 
      \leq \quad \Lambda ^2 _n  \quad + \quad 
      \bigg( \sum _\nu  m _\nu\bigg) ^2 
\ .  \label {eq:220} \end {equation}
The sector dependent mass scales $ \Lambda _n $ will not 
be needed below. They govern how much the particles can 
go off their equilibrium values
$\vec {\overline p} _{\!\bot}=\vec 0$ and 
$\overline x _\nu= m _\nu/ \sum _\nu  m _\nu$. 
The number of individual Fock states in DLCQ is thus finite,
and the Hamiltonian  matrix will have a finite dimension.
A finite matrix can be diagonalized numerically. 

This -- in a nutshell -- is the programme of Discretized Light-Cone 
Quantization \cite{pab85a}. 
One tends to conclude that the solution of the bound-state problem 
for a gauge field theory like QCD is solvable on a computer.  
\section {The Effective Interaction and the Tamm-Dancoff 
Approach} \label {sec:3}

DLCQ applied to gauge theory faces a formidable matrix 
diagonalization problem. 
The problem is even worse than in non-relativistic 
many-body theory. 
Here are the steps one should take in principle:  
In a first step one sets up the Fock space as 
discussed in the previous section. 
In a second step one calculates the finite dimensional  
Hamiltonian matrix as illustrated in Figure~\ref{fig:holy-1}. 
In a third step one diagonalizes the matrix 
numerically. Here is the bottleneck of the method:
Sooner or later, the matrix dimension exceeds
imagination, and other than in 1+1 dimensions 
one has to develop new tools by matter of principle.

Intuitively one aims at something like an effective interaction 
between a quark and an antiquark, similar to the effective 
interaction between a negative and a positive point charge.
Effective interactions are a well known tool in  many-body 
physics \cite{mof50}. To the community
the method is known as the Tamm-Dancoff-Approach, as 
applied first to Yukawa  theory for describing the 
nucleon-nucleon interaction \cite{tam45,dan50}. 
A look on its salient features is worth the effort.

Consider a Hamiltonian matrix as in Eq.(\ref{eq:314}) subject
to diagonalization,    $ H \vert \Psi \rangle = E \vert \Psi \rangle $. 
The matrix dimension be $N$. 
Explicitly written out, the eigenvalue equation reads 
\begin {equation} 
   \sum _{j=1} ^{N} \langle i \vert H \vert j \rangle 
                      \langle j \vert\Psi\rangle 
   =  E \ \langle i \vert\Psi\rangle 
\ , \ \quad{\rm for}\ i=1,2,\dots,N
. \label {eq:319} \end {equation}
The rows and columns of the matrix can always be split
into two parts. One speaks of  the $ P $-space  
$P = \sum _{j=1} ^n  \vert j \rangle\langle j \vert $
with $1<n<N$, and of the rest, 
the $Q$-space $Q\equiv 1-P$.
Eq.(\ref{eq:319}) can then be rewritten conveniently 
in terms  of {\em block matrices} like 
\begin {equation} 
  \pmatrix{ \langle P \vert H \vert P \rangle 
          & \langle P \vert H \vert Q \rangle \cr 
            \langle Q \vert H \vert P \rangle 
          & \langle Q \vert H \vert Q \rangle \cr} 
  \ \pmatrix{ \langle P \vert\Psi \rangle \cr 
              \langle Q \vert\Psi \rangle \cr  } 
  = E  
  \ \pmatrix{ \langle P \vert\Psi \rangle \cr 
              \langle Q \vert\Psi \rangle \cr  }
 \ , \label {eq:320} \end {equation}
or  explicitly
\begin {eqnarray} 
   \langle P \vert H \vert P \rangle\ \langle P \vert\Psi\rangle 
 + \langle P \vert H \vert Q \rangle\ \langle Q \vert\Psi\rangle 
 &=& E \:\langle P \vert \Psi \rangle 
 \ ,  \label{eq:321} \\   {\rm and} \qquad
   \langle Q \vert H \vert P \rangle\ \langle P \vert\Psi\rangle 
 + \langle Q \vert H \vert Q \rangle\ \langle Q \vert\Psi\rangle 
 &=& E \:\langle Q \vert \Psi \rangle 
\ . \label{eq:322}\end {eqnarray}
Rewrite the second equation as
\begin {equation} 
      \langle Q \vert E  -  H \vert Q \rangle 
    \ \langle Q \vert \Psi \rangle 
  =   \langle Q \vert H \vert P \rangle 
    \ \langle P \vert\Psi\rangle 
, \label {eq:326} \end {equation}
and observe that the quadratic matrix 
$ \langle Q\vert E -  H \vert Q \rangle $ could be inverted 
to express the Q-space wavefunction 
$\langle Q \vert\Psi\rangle $
in terms of the $ P $-space wavefunction
$\langle P \vert\Psi\rangle$. 
But here is the problem:   
The eigenvalue $ E $ is unknown at this point. 
One therefore solves first an other problem: One introduces
{\em the starting point energy} $\omega$ as a redundant
parameter  at disposal, and {\em defines the resolvent}
of the $ Q $-space matrix as the inverse of 
$\langle Q \vert\omega- H \vert Q \rangle$,  
\begin {equation} 
      G _ Q (\omega)   =  
      {1\over\langle Q \vert\omega- H \vert Q \rangle} 
\ . \label {eq:331} \end{equation} 
With  Eq.(\ref {eq:326} ) one {\em defines} then 
\begin {equation} 
         \langle Q \vert \Psi \rangle 
  \equiv
         \langle Q \vert \Psi (\omega)\rangle 
  =    G _ Q (\omega) \langle Q \vert H \vert P \rangle 
         \,\langle P \vert\Psi\rangle 
\ . \label {eq:332} \end {equation} 
Inserting this into Eq.(\ref{eq:321}) yields an eigenvalue
equation in the P-space
\begin {eqnarray} 
      \langle P \vert H _{\rm eff} (\omega ) 
      \vert P\rangle&&\langle P \vert\Psi_k(\omega)\rangle 
      = 
      E _k (\omega )\ \langle P \vert\Psi _k (\omega ) \rangle 
\ ,  \qquad\qquad{\rm with}  \label{eq:345} \\   
      \langle P \vert H _{\rm eff} (\omega) \vert P \rangle 
      &=& 
      \langle P \vert H \vert P \rangle +
      \langle P \vert H \vert Q \rangle 
      \ G _ Q (\omega) \ %
      \langle Q \vert H \vert P \rangle 
\ .  \label{eq:340} \end {eqnarray}
The effective interaction in the $ P $-space is thus well
defined: It is the  original matrix $ \langle P \vert H \vert P \rangle $
plus a part where the system is scattered virtually into the 
$ Q $-space, propagating there by impact of the true interaction, 
and finally scattered back into the $ P $-space. 
Every value of $\omega$ defines a different Hamiltonian 
and a different spectrum. 
Varying $\omega$ one generates a set of  
{\em energy functions} $ E _k(\omega) $. 
Whenever one finds a solution to the 
{\em fixpoint equation} \cite{pau81,zvb93}
\begin {equation}
E _k (\omega ) = \omega 
, \label {eq:350} \end {equation}
one has found one of the true eigenvalues and
eigenfunctions of $H$,  by construction. 

One should emphasize that one finds this way {\em all $N$ 
eigenvalues} of $ H $,  see also \cite{pau81}, irrespective of
how small one chooses the $ P $-space. 
Even if one chooses a $P$-space with 
`matrix' dimension $n=1$, the single energy function 
$E_1(\omega)$ contains the infomation on {\it all} 
eigenvalues:  The $N-1$ poles of  $E_1(\omega)$ generate
the $N$ eigenvalues by means of Eq.(\ref{eq:350}).
Explicit examples for that can be found 
in \cite{pau81,zvb93} and in Appendix~\ref{sec:c}. 
It looks as if one has mapped a difficult problem, the 
diagonalization of a huge matrix onto a simplier problem,  
the diagonalization of a much smaller matrix. 
This is true, but not completely. One has to invert a matrix.
In practice, the inversion of a matrix is as diffficult as its  
diagonalization. 
In addition,  one has to vary $\omega$ and solve the fixpoint 
equation (\ref{eq:350}).  The numerical work is thus rather 
larger than smaller as compared to a direct diagonalization. 

The advantage of working with the so defined effective
interaction is that resolvents can be approximated systematically. 
The two resolvents
\begin {equation}
     G _Q (\omega) =  {1\over \langle Q \vert 
           \omega - T - U  \vert Q \rangle} 
\ , \quad {\rm and\ } 
     G _0  (\omega) =  {1\over \langle Q \vert 
           \omega - T \vert Q \rangle} 
\ , \label {eq:352} \end {equation} 
defined once with and once without the non-diagonal 
interaction in the Hamiltonian $H = T + U$, respectively,
are identically related by 
\begin {eqnarray}   
          G _Q (\omega) 
&=& G _0  (\omega) + G_0 (\omega)  \ U \ G _Q (\omega)  
{\rm \ , \quad or\ by} \label {eq:353} \\   
     G _Q (\omega) 
 &=& G _0(\omega) + G_0 (\omega) U G _0 (\omega) 
    +    G_0 (\omega) U G_0 (\omega) U G _0 (\omega) 
    + \dots 
 \ ,  \label {eq:354} \end {eqnarray}
that is, by the {\em infinite series of perturbation theory}.
The point is, of course, that the kinetic energy $T$ is
a diagonal matrix which can be trivially inverted to get 
$G_0 (\omega)$.

After these formal considerations we return to gauge theory 
and its Fock space representations as displayed in 
Figure~\ref{fig:holy-1}. The identification of the
$q\bar q$-space with the $P$-space and the related 
effective interaction between one constituent ($q$)  with an
other ($\bar q$) appears as the most natural thing to do. 
But now the trouble starts: 
Like Tamm and Dancoff \cite{tam45,dan50}
one truncates the above series to the very first term
$H _{\rm eff} = H + H \vert Q\rangle \langle Q\vert 
  (\omega -T)^{-1} \vert Q\rangle  \langle Q\vert H  $.
But the Tamm-Dancoff  procedure fails for light-cone 
kinematics: When  taken literally, the effective interaction
has a non-integrable  singularity \cite{kpw92}.  
Even worse, simple estimates show that the series in 
Eq.(\ref{eq:354}) diverges \cite{kap92} order by order, 
all that irrespective of violating badly gauge invariance 
due to the truncation. One has to invent {\it ad hoc} 
measures to get things working \cite{kpw92}.  
One could try to {\em resum the series} in Eq.(\ref{eq:354}) 
at least partially, guided for example, like  in Quantum 
Electrodynamics,  by a smallness assumption on the
coupling constant.  When one tries that, one gets lost soon 
in the murky depths of perturbation theory 
{\em driven to all orders}, and quits the game. 
One the other hand, the procedure
displayed in Eqs.(\ref{eq:320}) to (\ref{eq:353}) 
{\em is in principle exact}.  Maybe one should look at it 
from an other point of view. One of the many conclusions
might be that the interaction term in the denominator  of
Eq.(\ref{eq:352})  should be dealt with in a better way, for
example,  by the `method of iterated resolvents' to be
presented next.

\section {The Method of Iterated Resolvents}
\label {sec:4}

DLCQ applied to gauge field theory is,  as mentioned, 
particular among matrix diagonalization problems to the 
extent that a {\em finite number} of  Fock-space sectors 
like in Figure~\ref{fig:holy-1} 
appears in the most natural way. 
The same holds true for quantum-mechanical $A$-body
problems where the Fock space can be classified in terms of
the {\em finite number} of  0-particle-0-hole (0-ph), 1-ph,
$\dots$, $A$-ph  sectors, see for example \cite{pau84}. 
In either case the block matrix in sparse.  Most of the
blocks are zero matrices due to the nature
of the underlying Hamiltonian. One should make use of that!

In DLCQ one is confronted with the diagonalization
of a {\em finite dimensional block matrix}, for which
Eq.(\ref{eq:320}) represents the example of  $2\times 2$ 
blocks. The step from Eq.(\ref{eq:320})  to Eq.(\ref{eq:345})
can then be interpreted as the reduction  of a block matrix
dimension  from 2 to 1. 
As a matter of fact, if one chooses the $Q$-space identical
with last sector $N_K$ in Figure~\ref{fig:holy-1} and the
$P$-space with the rest, $Q=1-P$, one reduces the block
matrix equation from dimension $N_K$ to $N_K-1$, with an
effective interaction acting in the now smaller space. 
This procedure can be repeated, and again iterated, until one
arrives at an effective interaction of block matrix dimension
$n=1$, {\em with a well defined} effective interaction in the
`1'-space. All that is needed is a reasonable notation
\cite{pau93}. 

Suppose, in the course of this reduction, one has 
arrived at block matrix dimension $n$, 
with $1<n<N_K$.  Denote the corresponding effective interaction  
$H_n (\omega)$. The eigenvalue problem corresponding to 
Eq.(\ref{eq:345}) reads now
\begin {equation} 
   \sum _{j=1} ^{n} \langle i \vert H _n (\omega)\vert j \rangle 
                      \langle j \vert\Psi  (\omega)\rangle 
   =  E (\omega)\ \langle i \vert\Psi (\omega)\rangle 
\ , \ \quad{\rm for}\ i=1,2,\dots,n
. \label {eq:406} \end {equation}
Observe that $i$ and $j$ refer here to sector numbers.  
Now,  in analogy to Eqs.(\ref{eq:331}) and
(\ref{eq:332}), define
\begin {eqnarray} 
            G _ n (\omega)   
&=&   {1\over \langle n \vert\omega- H \vert n \rangle} 
\ , \qquad {\rm and} \label {eq:410} \\  
            \langle n \vert \Psi (\omega)\rangle   
&=&   G _ n (\omega) 
   \sum _{j=1} ^{n-1} \langle n \vert H _n (\omega)\vert j \rangle 
   \ \langle j \vert \Psi (\omega) \rangle 
\ , \label {eq:411} \end {eqnarray}
respectively. The effective interaction 
in the  ($n -1$)-space becomes then \cite{pau93} 
\begin {equation}  
       H _{n -1} (\omega) =  H _n (\omega)
  +  H _n(\omega) G _ n  (\omega) H _n (\omega)
\label {eq:414} \end {equation}
for every block matrix element 
$\langle i \vert H _{n-1}(\omega)\vert j \rangle$.  
To get the corresponding eigenvalue equation one
substitutes $n$ by $n-1$ everywhere in 
Eq.(\ref{eq:406}). Everything proceeds like in
section~\ref{sec:3},   including the fixed point equation  
$ E  (\omega ) = \omega $.
But one has achieved much more: Eq.(\ref{eq:414}) is a 
{\em recursion relation} which holds for all $1<n<N_K$!
The only convention one must have, of course, is that one 
has started from the bare interaction $H$, thus
$H_{N_K}=H$. The rest is algebra and interpretation.

How can the procedure be interpreted? One gets some
inspiration from the paradigm of a $4\times 4$ 
block matrix as explicitly dealt with in Appendix~\ref{sec:c}. 
The consequences of Eq.(\ref{eq:414}) are worked out
in Appendix~\ref{sec:d}: When expressed in terms of the
bare interaction (matrices) $H$ and the resolvents $G_n$, 
the effective interaction in sector $n$  
wants to develope {\em chains} like 
\begin {equation}  
       H _n (\omega) =  \dots\  + H G _ l (\omega) H 
       G _ m(\omega) H G _ r (\omega) H +\ \dots
\ , \qquad{\rm\ with\ } l,m,r > n 
. \label {eq:4015} \end {equation}
The order of the chains is given by the number of 
propagators and not, as usual, by the power of
the coupling constant. The maximum order of the chains 
{\em is strictly finite for a finite matrix}, as opposed to the 
infinite series of perturbation theory.
The sector numbers in the chains  
are not arbitrary, but obey the rules
tabulated in Appendix~\ref{sec:d}. For example,
the effective interaction $H_n$ has no propagator with
sector number $n$. This implies that the system {\em can 
not fall back}, a feature distincly different from the usual
perturbative propagators displayed in Eq.(\ref{eq:352}). 
Moreover, most of the chains vanish from the outset for the
DLCQ-Hamiltonian:  It suffices that only one of the four 
bare block matrix elements appearing in Eq.(\ref{eq:4015}), 
namely $\langle n\vert H \vert l \rangle $, 
$\langle l\vert H \vert m \rangle $, 
$\langle m\vert H \vert r \rangle $, or  
$\langle r\vert H \vert n \rangle $  is a zero-matrix.
Unpleasant is to some extent, that one has to deal with
`resolvents of resolvents',  or with `iterated resolvents'. 
On the other hand, each
resolvents represents a resummation of 
all orders in perturbation theory, and it is this
particular aspect  which makes them interesting, at least.

\section {Application to Quantum Chromodynamics } 
\label{sec:6}

The method of iterated resolvents is almost ideally suited for
dealing with the DLCQ matrix for Quantum Chromodynamics,  
see Figure~\ref{fig:holy-1}, and for deriving the 
effective Hamiltonian in the $q\bar q$-space. 

In doing so, we actually shall use a trick which will simplify 
considerations enormously. 
Practitioners in Light-Cone Time-Ordered Perturbation 
Theory \cite{leb80,leb81,brp91,tab91} know that they can
omit  the instantaneous interactions $F$ and $S$ until 
{\em they actually compute} a particular diagram. 
Then, every intrinsic line in a graph must be combined with 
the instantaneous partner  line associated  with the gauge artefacts. 
Only then, the sum of all time ordered diagrams becomes manifestly
identical with gauge-invariant Feynmann scattering amplitudes.
There is no exception known to this rule, thus far, in all graphs
computed explicitly.
In the sequel, this `gauge trick' \cite{tab91} is adopted by setting 
formally to zero all block matrices in Figure~\ref{fig:holy-1}
which are  gauge remnants. One then gets a block matrix 
structure as displayed in Figure~\ref{fig:holy-2}. The 
extreme sparseness of this block matrix will allow us to 
carry the procedures through to the end. 
When eventually arriving there, we will check explicitely 
in Appendix~\ref{sec:f} whether  the restitution of the gauge 
artefacts changes anything in the structure 
of the solution.  The answer will be: `It doesn't'.

\def\d{$\bullet$}  \def\v{ V }  \def\b{ $\cdot$ } \def\s{ }   \def\f{ }
\begin{figure} [t]
\begin {center}
\caption{\label{fig:holy-2} \sl
    The Hamiltonian matrix for a SU(N)-meson when only 
     vertex diagrams ($V$) are included.
     Com\-pare also with Figure~\protect\ref{fig:holy-1}.
     Zero matrices are marked by ($\cdot$). 
     (In the preprint this table is replaced
     by a figure with the diagrams.)
}\vspace{1em}
\begin {tabular}  {||cc||cc||c|ccc|cccc|ccccc||}
\hline \hline 
     &   &        & N$_p$ & 
     2 & 2 & 3 & 4 & 3 & 4 & 5 & 6 & 4 & 5 & 6 & 7 & 8 
\\ 
   K & N$_p$ & Sector & n & 
     1 & 2 & 3 & 4 & 5 & 6 & 7 & 8 & 9 &10 &11 &12 &13 
\\ \hline \hline   
    1 &  2 & $ q\bar q\, $ &  1 & 
    \d &\s &\v &\f &\b &\f &\b &\b &\b &\b &\b &\b &\b 
\\ \hline
    2 &  2 & $ g\ g\ $ &  2 & 
    \s &\d &\v &\b &\v &\f &\b &\b &\f &\b &\b &\b &\b 
\\
    2 &  3 & $ q\bar q\, \ g $ &  3 & 
    \v &\v &\d &\v &\s &\v &\f &\b &\b &\f &\b &\b &\b 
\\
    2 &  4 & $ q\bar q\, q\bar q\, $ &  4 & 
    \f &\b &\v &\d &\b &\s &\v &\f &\b &\b &\f &\b &\b 
\\ \hline
    3 &  3 & $ g \ g \ g $ &  5 & 
    \b &\v &\s &\b &\d &\v &\b &\b &\v &\f &\b &\b &\b 
\\
    3 &  4 & $ q\bar q\, \ g \ g $ &  6 & 
    \f &\f &\v &\s &\v &\d &\v &\b &\s &\v &\f &\b &\b 
\\
    3 &  5 & $ q\bar q\, q\bar q\, \ g $ &  7 & 
    \b &\b &\f &\v &\b &\v &\d &\v &\b &\s &\v &\f &\b 
\\
    3 &  6 & $ q\bar q\, q\bar q\, q\bar q\, $ &  8 & 
    \b &\b &\b &\f &\b &\b &\v &\d &\b &\b &\s &\v &\f 
\\ \hline
    4 &  4 & $ g \ g \ g \ g $ &  9 & 
    \b &\f &\b &\b &\v &\s &\b &\b &\d &\v &\b &\b &\b 
\\
    4 &  5 & $ q\bar q\, \ g \ g \ g $ & 10 & 
    \b &\b &\f &\b &\f &\v &\s &\b &\v &\d &\v &\b &\b 
\\
    4 &  6 & $ q\bar q\, q\bar q\, \ g \ g $ & 11 & 
    \b &\b &\b &\f &\b &\f &\v &\s &\b &\v &\d &\v &\b 
\\
    4 &  7 & $ q\bar q\, q\bar q\, q\bar q\, \ g $ & 12 & 
    \b &\b &\b &\b &\b &\b &\f &\v &\b &\b &\v &\d &\v 
\\
    4 &  8 & $ q\bar q\, q\bar q\, q\bar q\, q\bar q\, $ & 13 & 
    \b &\b &\b &\b &\b &\b &\b &\f &\b &\b &\b &\v &\d 
\\ \hline\hline
\end {tabular}
\end {center}
\end{figure}

When writing down the chains for the block matrix displayed in 
Figure~\ref{fig:holy-2} according to the rules of the method of
iterated resolvents,  the effective Hamiltonian in the 
$q\bar q$-space becomes
\begin {equation}  
     H _1 =  T_1 + V G _3 V + V G _3 V  G _2 V G _3 V 
\ . \label {eq:610} \end {equation} 
Only two chains survive the procedure, one with one, and
one with three propagators $G_j=G_j(\omega)$.
One checks explicitly with Eqs.(\ref{eq:442})  and  
(\ref{eq:444}) that  no chains  with four, five or more propagators, 
are possible, even not for the case of arbitrarily large harmonic
resolution $K$. Eq.(\ref{eq:610}) stands thus for full QCD in the
continuum limit. The gauge trick is thus a great help: Including 
the gauge artefacts, the effective Hamiltonian in the 
$q\bar q$-space would given 32 non-trivial chains up to 
order 4. These are explicitly tabulated in Appendix~\ref{sec:d}.

Why are there only two chains in Eq.(\ref{eq:610})?
The system can scatter out of the $q\bar q$-space 
into the  $q\bar q\,g$-space by emitting a gluon from  
the quark or from the antiquark. There is no other way since
all other block matrix elements $\langle 1\vert H \vert j\rangle$ 
vanish according to Figure~\ref{fig:holy-2}. 
This kills all possible chains which do not start like 
$V G _3 V  G _j V G _l \dots\ $. The scattering from the
$q\bar q\,g$-space proceeds either by scattering back 
to the $q\bar q$-space, which is the first chain in
Eq.(\ref{eq:610}), or by scattering into another space 
by a sequence like $G_3 V G_j$. 
The rules of MIR require $ 3>j>1$.  The only solution  $j=2$ 
gives the second chain. 
\begin{figure} [t]
\begin{minipage}[t]{80mm} \makebox[0mm]{}
\epsfxsize=80mm\epsfbox{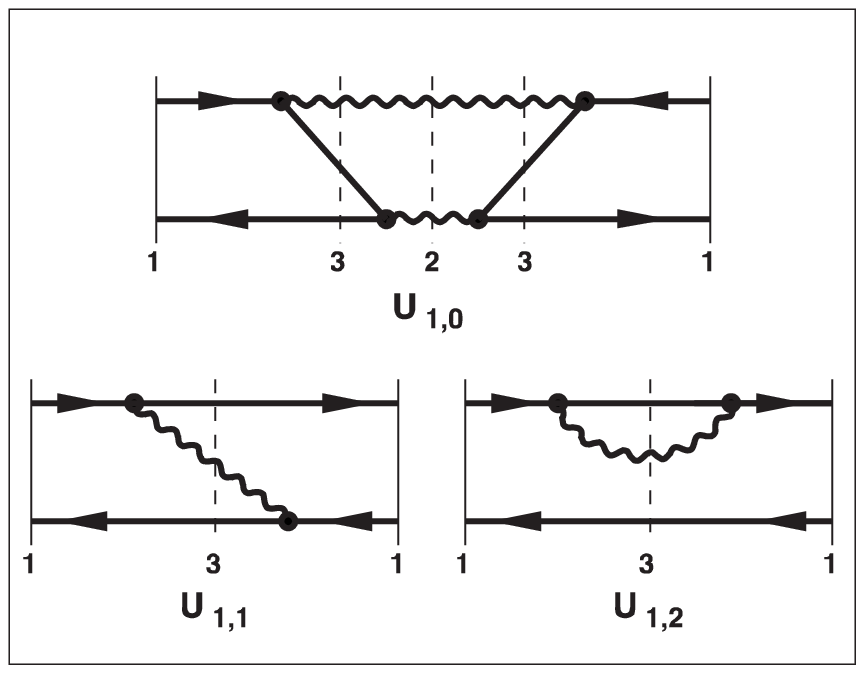}
\caption{\label{fig:6_1} \sl
  Three graphs of the effective interaction in the 
  $q\bar q$-space.~---
  The lower two graphs correspond to the chain $VG_3V$,
    the upper corresponds to $VG_3V G_2V G_3V$. 
   Propagators are represented by vertical dashed lines 
   with a subscript `$n$' for the sector. 
} \vfill \end{minipage}
\hfill
\begin{minipage}[t]{75mm} \makebox[0mm]{}
\makebox[0mm]{}
\epsfxsize=80mm\epsfbox{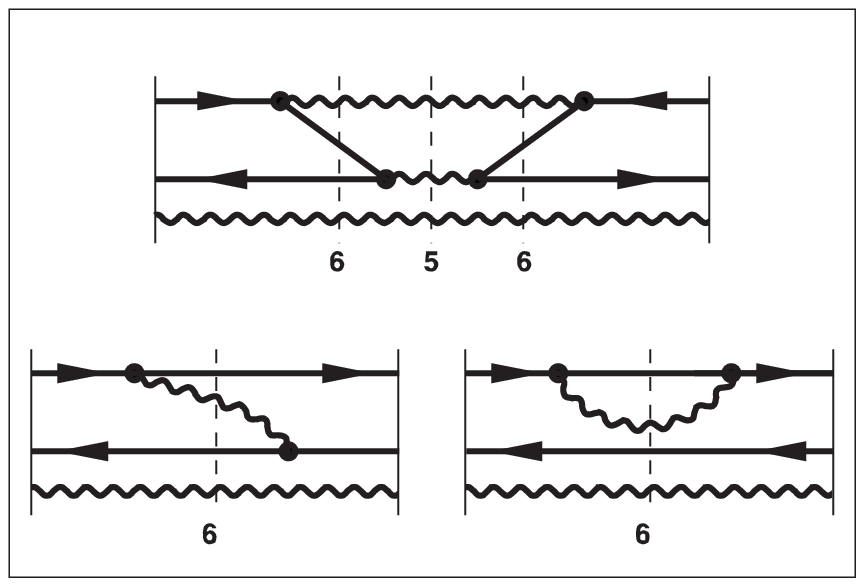}
\caption{\label{fig:6_2} \sl
    Graphs of the  spectator interaction 
    in the $q\bar q\,g$-space. 
    Note the role of the gluon  as a spec\-tator.
}\vfill \end{minipage}
\end{figure}

What do the chains in Eq.(\ref{eq:610}) mean in terms 
of physics? We find it convenient to illustrate that by the 
diagrams in Figure~\ref{fig:6_1}. 
The first term in the effective Hamiltonian $H_1$
refering  to the kinetic energy ($T_1$) in the 
$q\bar q$-space needs no illustration.   In the second term
($V G _3 V$) the vertex interaction $V$ creates a gluon and 
scatters the system  from the  $q\bar q(1)$-sector into the
$q\bar q\,g$-space. As
indicated by the vertical line with the subscript `3', the three
particles propagate there in `3'-space under impact of the full
interaction before the gluon is annihilated. 
The gluon can be absorbed either by the antiquark
or by the quark, corresponding to
the two `interactions' $U_{1,1}$ and $U_{1,2}$, respectively,
in the figure. Two further graphs, describing the
irradiation of the gluon by the antiquark, are not shown; 
like throughout in the sequel we use the graphs for the
purpose of  illustration rather then of a complete description. 
As the inverse of the non-diagonal Hamiltonian $H_3(\omega)$, 
the propagator $G _3 (\omega)$ is non-diagonal as well, 
in general. Therefore, despite the total (space-like) momenta
being conserved strictly by the interaction and thus by  
propagation, the space-like momenta of the individual particles 
can change.
A finite size box instead of a thin line would thus be more appropriate
for graphically representing propagators. With this proviso
in mind, and the recognition that an interaction can have
several graphs, we proceed. The last term, finally, in 
Eq.(\ref{eq:610}) is represented correspondingly by the graph
$U_{1,0}$ in Figure~\ref{fig:6_1}. 

\subsection {The quark-pair-glue propagators,
   and propagation `in medium'} \label {sec:6.2}

The mind trained by perturbation theory wonders about
all  those zillions of graphs, he is used to work with. 
The answer is that they have neither been ommitted nor 
neclected but that they reside  in the propagators $G _i$, 
very effectively `resumed to all orders'. How do they look?

\begin{figure} [t]
\begin{minipage}[t]{80mm} 
\makebox[0mm]{}
\epsfxsize=80mm\epsfbox{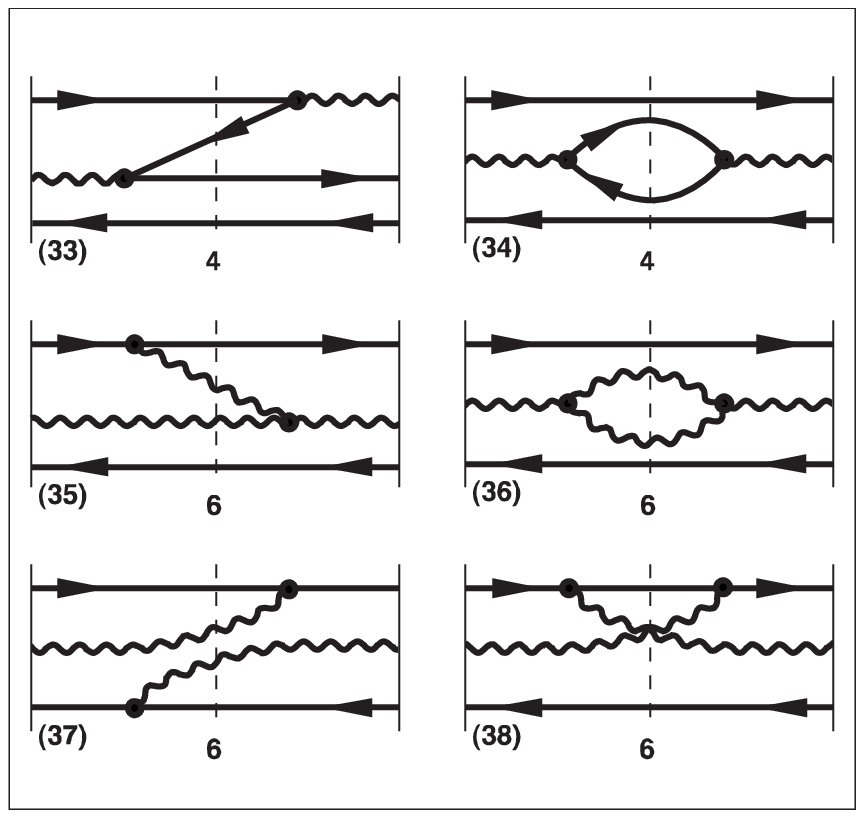}
\caption{\label{fig:6_3} \sl
    Six graphs of the  participant interaction 
    in the $q\bar q\,g$-space.
} \vfill \end{minipage}
\hfill
\begin{minipage}[t]{70mm} 
\makebox[0mm]{}
\epsfxsize=70mm\epsfbox{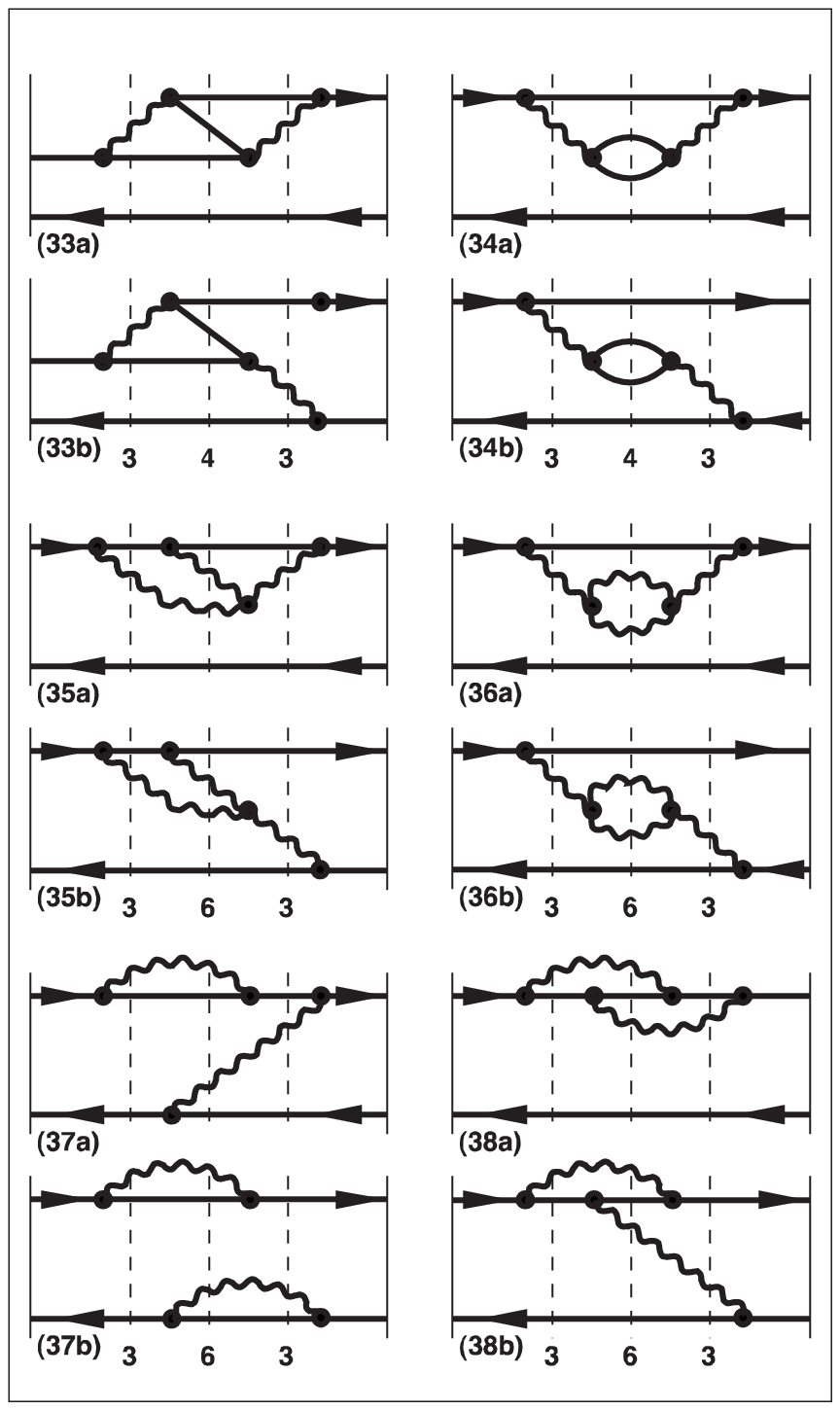}
\caption{\label{fig:6_4} \sl
   Twelve graphs corresponding to Fig.~\protect\ref{fig:6_3} 
    with the gluons absorbed after pro\-pa\-gation.
}  \vfill  \vfill \end{minipage}
\end{figure}

By the rules established in Appendix~\ref{sec:d}, one
obtains consecutively for the effective Hamiltonians in the 
sectors with one $q\bar q$-pair and 1,2,3 or more gluons 
\begin {eqnarray} 
     H _3 
&=&  T_3 + V G _6 V + V G _6 V  G _5 V G _6 V + V G _4 V 
\ , \label {eq:620}  \\  
     H _6 
&=&   T_6 + V G _ {10} V+ V G _{10} V  G _9 V G _ {10} V 
                 + V G _7 V 
\ , \label {eq:621} \\
     H _{10} 
&=&  T_{10} + V G _{15} V + V G _{15} V  G _{14} V G _{15}V
      +V G _{11} V  
\ , \label {eq:622} \end {eqnarray}
respectively, to whose resolvents we refer collectively as the 
quark-pair-glue propagators. 
Note that these are all exact relations, by the same reasons as 
discussed above, for the $q\bar q(1)$-sector.
Note also that they all have the same structure. 
They differ from the effective Hamiltonian in the 
$q\bar q$-space, Eq.(\ref{eq:610}), only by their
respective last term, in which a gluon is annihilated into a
$q\bar q$-pair. Since the $q\bar q$-sector has no gluon,
by definition, the corresponding term must be absent there.
In passing, one notes the `replica-structure' of these
Hamiltonians and propagators. Like in Russian puppets
one has propagators of propagators, and a certain amount of 
self-similarity cannot be denied. These strucures seem to be
particular to gauge theory, abelian or not. 

Let us have a look at the diagrammatic representation of 
these Hamiltonians, and in particular on the effective
Hamiltonian in the $q\bar q\,g(3)$-sector, Eq.(\ref{eq:620}).
A typical collection is given in Figures~\ref{fig:6_2} and 
\ref{fig:6_3}. The subset of graphs displayed in 
Figure~\ref{fig:6_2} looks exactly like the effective interaction 
in Figure~\ref{fig:6_1}, except of the additional gluon. 
The gluon does not change quantum numbers under impact
of the interaction, and acts like a spectator. We thus refer to 
the graphs in Figure~\ref{fig:6_2} as the  
`spectator interaction' $\overline U _3$. In the graphs of 
Figure~\ref{fig:6_3}, the gluon undergoes scatterings 
which correspondingly are refered to as the 
`participant interaction' $\widetilde U _3$, in the sequel. 
The effective Hamiltonians the quark-pair-glue sectors 
$q\bar q\,g\dots g$, as given in Eqs.(\ref{eq:620})
to (\ref{eq:622}), can thus be written  
 \begin {equation}  
     H _n =  T  _n + \overline U _n + \widetilde U _n
\ , \quad{\rm for}\quad  n= 3,6,10,15,\dots
\ ,  \label {eq:626} \end {equation} 
collectively, and with the obvious corresponding definitions. 

In terms of physics, the separation into spectators and
participants becomes more transparent 
in a perturbative consideration.
Think of the incoming gluon in Figure~\ref{fig:6_3} as being
irradiated by a quark, and of the outgoing gluon as to be
absorbed either by the quark (a) or by  the antiquark (b),
and draw the corresponding diagrams. Each diagram in
Figure~\ref{fig:6_3} will then be associated with two
diagrams in Figure~\ref{fig:6_4}. 
The labels on the diagrams in the figures have no other 
purpose in the present context than to help the reader with 
this association.  The diagrams in Figure~\ref{fig:6_4}
look familiar. Would one do plain perturbation theory,
they would be diagrams of order 4 in the coupling constant.
Indeed, they look like  the familiar mass, vertex or vacuum 
`corrections', which are associated usually with the 
{\em running coupling constant} \cite{grw73,pol73}. 
Here, however, we continue to deal with propagators 
in medium and take the analogue only as an argument 
why the separation into `spectators' and `participants' 
might be useful. 

The essence is that the propagators $\overline G _n$
associated  with the spectator interaction $\overline U _n$, {\it i.e.}
\begin {equation} 
     \overline G _n  =   {1 \over  \omega - T _n - \overline U _n }
\ , \qquad \biggl( {\rm while\ } G _n   =
     {1 \over  \omega - T _n - \overline U _n - \widetilde U _n }
      \equiv {1\over \omega - H _n} \ \biggr)
     \ , \label {eq:641} \end {equation}  
are identically related to the full propagators $G_n$ by 
\begin  {equation} 
     G _n    =   \overline G _n   +  \overline G _n \,
                        \widetilde U _n \, G _n         
\ . \label {eq:642} \end {equation} 
As usual, they can be written as an infinite series
\begin  {equation} 
     G _n  = \overline G _n
     + \overline G _n \widetilde U _n\, \overline G _n 	
     + \overline G _n \widetilde U _n\, \overline G _n 
                                     \widetilde U _n\, \overline G _n   + \dots
    \ . \label {eq:644} \end  {equation} 
The main difference to the usual series like in 
Eq.(\ref{eq:353}) is, that there the `unperturbed propagator'
$ G _0 (\omega) $ refers to the system without interactions
while here the `unperturbed propagators' 
$\overline G _n$ contain the
interaction in the well defined form of $\overline U_n$.
One therefore deals here with `perturbation theory in medium'.
Note that the present series is different 
from the above Eq.(\ref{eq:353}) also with respect to the
physics: The system stays in sector $n$. This allows for an
identical rearrangement of the series
\begin  {equation} 
      G _n 
     =  \left[  1  
     + {1\over 2} \overline G _n \widetilde U _n 
     + {3\over 8} \overline G _n \widetilde U _n \,  
                             \overline G _n \widetilde U _n  + \dots
  \right]  \overline G _n \left[ 1 
     + {1\over 2} \widetilde U _n\, \overline G _n
     + {3\over 8} \widetilde U _n\, \overline G _n 
                             \widetilde U _n\, \overline G _n + \dots\right]
\ , \label {eq:648} \end {equation} 
which can be verified order by order and which, 
to our recollection, has not been given before.
The series in the square bracket are known to be 
the inverse square, {\it i.e.}
\begin  {equation}
       R  _n
  =  {1\over \sqrt { 1 - \overline G _n \widetilde U _n}  }
  =   1
  +  {1\over 2} \overline G _n \widetilde U _n
  +  {3\over 8} \overline G _n \widetilde U _n   \,
                           \overline G _n \widetilde U _n + \dots  
\ ,  \label {eq:6510} \end {equation} 
and therefore the full propagators can be 
rewritten identically as 
\begin  {equation} 
      G _n = R _n^\dagger \overline G _n R _n
=  {1\over \sqrt {1 - \widetilde U _n \overline G _n} }\ \overline G _n
\   {1\over \sqrt {1 - \overline G _n \widetilde U _n} }
\ . \label {eq:6520} \end {equation} 
This result can be verified also by closed operator relations. 
First, one notes the strict identity
\begin  {equation} 
  (1-\widetilde U _n \overline G _n )(\omega - \overline H _n ) =
  \left( 1-\widetilde U _n {1\over \omega - \overline H _n}
  \right) (\omega - \overline H _n ) = \omega - H _n 
\ . \label {eq:6530} \end {equation} 
Since its adjoint is also correct, one gets
\begin  {equation} 
      G _n ^\dagger G _n 
=  {1\over 1-U _n\overline G _n} {1\over(\omega -H _n)^2}
     {1\over 1-\overline G _n U_n}
\ . \label {eq:6540} \end {equation} 
This is a positive valued operator. Taking the square root 
with this precaution, gives Eq.(\ref{eq:6520}) as an identity.
One thus can proceed without having to worry on the square
roots, their  ambiguities of signs, and on the convergence
of the  series. The square matrix $ R $ will always 
be sandwiched between a quark-pair-glue propagator
$\overline G$ and two vertex interactions $V$, for which
reason  it is convenient to introduce $\overline V$ as an 
abbreviation, defined by
\begin  {equation} 
     V\,G _n (\omega)\, V 
=        V  \,R_n^\dagger(\omega)\overline G_n(\omega)
R_n(\omega)\,V      
\equiv \overline V \,\overline G_n\,\overline V      
\ . \label {eq:652} \end {equation} 
Below,  a very natural and physical interpretation is given to
 the operator $R$,  as being related to 
`running coupling constant', 
but here we continue to proceed formally.  
We use the above findings to systematically rewrite 
Eqs.(\ref{eq:610})-(\ref{eq:622}). 
Working upwards in the hierarchy, one gets consecutively: 
\begin {eqnarray} 
     \overline H _6 
&=&   T_6 + \overline V \,\overline G _ {10} \overline V
                     + \overline V \,\overline G _{10} \overline V  
              G _9 \overline V \,\overline G _ {10} \overline V 
\ , \label {eq:662} \\
     \overline H _3 
&=&  T_3 + \overline V \,\overline G _6 \overline V 
                    + \overline V \,\overline G _6 \overline V  
              G _5\overline V \,\overline G _6 \overline V 
\ , \label {eq:663} \\      
       H _1= \overline H _1
&=&  T_1 + \overline V \,\overline G _3 \overline V 
                    + \overline V \,\overline G _3 \overline V  
             G _2 \overline V \,\overline G _3 \overline V 
\ . \label {eq:664} \end {eqnarray}
Instead of being similar, the sector Hamiltonians 
$\overline H _n = T_n + \overline U _n$ 
become equal  for suffiently large $K$. 
All what is different is that they act in different 
quark-pair-glue spaces, namely 
in the $q\bar q$-spaces with 1, 2 or more gluons. 

The gluons, however, by construction do not take part in the 
interaction. They act like inert spectators, very much like 
displayed already in Figures~\ref{fig:6_1} and~\ref{fig:6_2}
for the interactions in the $q\bar q(1)$- and the 
$q\bar q\,g$-space, respectively. 
All what one has to do is to update the figures and to replace 
{\em each point-like vertex} by  say a square, which should
symbolize  the impact of the operator $R$.

\subsection{The key point: The quark-pair-glue
resolvents `in the solution'}   
\label{sec:e}  

The operators  $\overline H_n = \overline H_n (\omega)$ 
are  {\it bona fide} Hamiltonians which can be diagonalized on
their own 
\begin  {equation}
       \overline H_n (\omega)  \vert \Psi _n (\omega) \rangle
       = E_{n} (\omega) \vert \Psi _n (\omega) \rangle
\ , \quad{\rm for}\quad  n= 1,3,6,10,\dots
\ .  \end {equation} 
Since the (spectator) gluons do not interact with the bound 
$q\bar q$ system, by construction, one can relate the various
spectra, knowing nothing on $\overline U_n$ except that the 
same potential acts only between the quarks. 
For $n=1$, the Fock states are $\vert q;\bar q\rangle 
   = b^\dagger _q d^\dagger_{\bar q} \vert vac \rangle $
and the eigenvalue equation reads
\begin  {equation}
       \sum _{q^\prime,\bar q^\prime}
       \langle q;\bar q\vert H_{1}(\omega)
       \vert q^\prime;\bar q^\prime\rangle
       \langle q^\prime;\bar q^\prime\vert
       \Psi_b(\omega) \rangle =  M_{b} ^2(\omega) 
       \langle q;\bar q\vert\Psi_b(\omega) \rangle
\ .   \end {equation} 
The eigenstates labeled by $b$ form a complete set of states
and  denumerate the physical mesons and their spectrum. 
They correspond to physical particles up to the
fact that their invariant mass  $M_{b} = M_{b} (\omega) $ 
still depends on the starting point energy.
Correspondingly for $n=3$, one has 
$\vert q;\bar q;g \rangle = b^\dagger_q d^\dagger_{\bar q}
   a^\dagger _g \vert vac \rangle $ and
\begin  {equation}
       \sum _{q^\prime,\bar q^\prime,g^\prime}
       \langle q;\bar q;g\vert
       \overline H_{3} (\omega) \vert q^\prime;\bar q^\prime
       ;g^\prime\rangle\langle q^\prime;\bar q^\prime
       ;g^\prime\vert \psi_{b,s}(\omega) \rangle
       = M_{b,s} ^2(\omega) 
       \langle q;\bar q;g\vert\psi_{b,s}(\omega) \rangle
\ .  \end {equation} 
Because the gluon is a {\em non-interacting spectator}, 
the eigenfunction is a product state
$\vert \psi_{b,s}(\omega) \rangle
   =\vert \Psi_{b}(\omega) \rangle
   \otimes \vert \varphi_{s}\rangle$, 
labeled by the {\em two indices} $b$ and $s$. In fact, the
eigenvalues in this sector {\it must be}
\begin  {equation}
      M_{b,s} ^2
      = {M_b^2 + \vec k_{g_{\!\perp}}^{\,2} \over (1-x_g)} 
      + {\vec k_{g_{\!\perp}}^{\,2} \over x_g} 
\  ,   \end {equation}
corresponding to the free gluon moving with momentum
(fraction)  $\vec k_{g_{\!\perp}}\, (x_g)$ relative to the meson 
with mass $M_b$ and momentum (fraction) 
$\vec k_{b_{\!\perp}}= -\vec k _{g_{\!\perp}}\ (x_b=1-x_g)$. 
Correspondingly the eigenvalue in the $q\bar q\,g\dots g$
sectors are  
\begin  {equation}
      M_{b,g\dots g} ^2 =
      {M_b^2 + \vec k_{g_{\!\perp}}^{\,2} \over (1-x_g)} +
      {\mu^2_{g\dots g} + \vec k_{g_{\!\perp}}^{\,2} \over x_g} 
\  .   \end {equation}
Here the meson moves against the cluster of free gluons
with free invariant mass $\mu_{g\dots g} $. Note that these
statements are {\it frame independent} and  a direct
consequence of the front form. Only in the front form 
the transition to a moving frame is trivial. In the the instant 
form the boost operators are  non-diagonal and complicated.
Note also that these statements hold {\it if and only if} the
interaction in  all sectors is identical, {\it i.e.}
\begin  {equation}
     U_n (\omega) = U_1 (\omega)
\ .  \end {equation} 
Is this really true? A word of caution seems to be in place.
Actually, {\it it is not true for any finite $K$}. 
The argument is simple: The `continued fraction' 
behind the resolvents of resolvents and the Russian puppet 
structure of the theory becomes interrupted in the last sector.
For a finite $K$ there is always a last one, no matter how
large $K$ is chosen, and thus the continued fractions in the 
different sectors do not have the same length and therefore 
are different. In the {\it continuum limit}, however, this
argument of denumeration does not hold. Thus, for any finite
but  sufficiently large $K$ the above statement is 
only `sufficiently true'.

The resolvents of $\overline H _n$,  in general, are
non-diagonal matrices in Fock space representation.
Transforming to the representation in which 
$\overline H _n$ is diagonal, one gets for $n=3$
\begin  {eqnarray}
      \langle q;\bar q;g \vert \overline G _3 (\omega)
      \vert q ^\prime;\bar q^\prime;g ^\prime \rangle 
      &=& \langle q;\bar q;g \vert  
      {1\over \omega - \overline H _3(\omega)} 
      \vert q ^\prime;\bar q^\prime;g ^\prime \rangle 
\nonumber\\
      &=& \sum _{b,s} \displaystyle
      \langle q;\bar q; g 
      \vert\Psi_{b,s}\rangle\langle \Psi_{b,s}\vert  \,{1
      \over  \omega -
      {M_b^2 + \vec k_{g_{\!\perp}}^{\,2} \over(1-x_g)}  - 
      {\vec k_{g_{\!\perp}}^{\,2} \over x_g} }\, 
      \vert\Psi_{b,s}\rangle\langle \Psi_{b,s} 
      \vert q^\prime;\bar q ^\prime; g^\prime \rangle       
\ .  \end {eqnarray}
Consider now  the resolvent  $\overline G(\omega)$ as function 
of $\omega$. It depends, of course, on the value of $\omega$ 
{\em and  on the spectral distribution of the eigenvalues}. 
Henceforward, as being reasonable for a `bound state
calculation',  we shall {\em restrict to the spectral region where 
the eigenvalues are sufficiently well separated from each other}.   
$\overline G(\omega)$  will be strongly peaked whenever  
$ \omega $ hits one of the true eigenvalues, say
$\omega = M^2_{b^\prime} (\omega) $, 
according to Eq.(\ref{eq:350}).
{\em At the peak} of this function, however, the following
sequence of steps can be performed: 
\begin  {eqnarray}
      \langle q;\bar q;g \vert \overline G _3 (\omega)
      \vert q ^\prime;\bar q^\prime;g ^\prime \rangle 
      &=& \sum _{b,s} \displaystyle
      \langle q;\bar q; g 
      \vert\Psi_{b,s}\rangle{1\over M^2_{b^\prime} -
      {M_b^2 + \vec k_{g_{\!\perp}}^{\,2} 
      \over (1-x_g)}  -
      {\vec k_{g_{\!\perp}}^{\,2}\over x_g}  }
      \langle \Psi_{b,s} 
      \vert q^\prime;\bar q ^\prime; g^\prime \rangle       
\label{eq:57} \\
      &\simeq& \sum _{b,s} \displaystyle
      \langle q;\bar q; g 
      \vert\Psi_{b,s}\rangle{1\over 
      M^2_{b^\prime} - M_b^2 - 
      {\vec k_{g_{\!\perp}}^{\,2} \over x_g}} 
      \langle \Psi_{b,s} 
      \vert q^\prime;\bar q ^\prime; g^\prime \rangle       
\label{eq:58} \\
      &\simeq& -{x_g\over \vec k_{g_{\!\perp}}^{\,2} }
      \sum _{b,s} \displaystyle
      \langle q;\bar q; g \vert\Psi_{b,s}\rangle
      \langle \Psi_{b,s} 
      \vert q^\prime;\bar q ^\prime; g^\prime \rangle       
\label{eq:59} \\
      &=& - {x_g\over \vec k_{g_{\!\perp}}^{\,2} }
      \ \langle q;\bar q; g 
      \vert q^\prime;\bar q ^\prime; g^\prime \rangle       
\ .  \end {eqnarray}
The first step invokes a {\em new smallness condition}:
\begin  {equation}
      x_g \ll \epsilon \leq 1 \qquad{\bf and}\quad
      \vec k _{g_{\!\perp}} ^2 \ll \widetilde\Lambda ^2\leq M_b^2 
\ . \label{eq:e80}\end {equation}
It has a completely different origin than the cut-offs
$\Lambda_n$  introduced for Fock-space regularization. 
The second step, including closure,  is not only 
permitted but also a valid approximation in line with 
standard many-body technology.

No approximation, however, is involved relating the off-shell 
mass $\vec k _{g_{\!\perp}} ^{\,2} /x_g$ of the gluon to the 
other particles in $\vert q;\bar q;g\rangle$. Denoting the
single-particle four-momentum of the gluon, the quark, and
the antiquark by $k_g$,$k_q$, and $k_{\bar q}^\prime$,
respectively, only well-known light-cone kinematics is needed
to establish that the two four-vectors  
\begin{equation}      
       l_{\bar q} ^\mu = \left(k_g + k_{\bar q}^\prime 
       - k_{\bar q}\right) ^\mu
       = {\vec k _{g_{\!\perp}} ^{\,2}\over  2k_g^+}\ \eta ^\mu
       \quad{\rm and}\quad
       l^\mu_q = \left(k_g + k_q - k_q^\prime\right) ^\mu
       = {\vec k _{g_{\!\perp}} ^{\,2}\over  2k_g^+}\ \eta ^\mu
\end{equation}
are identical, and proportional to the time-like vector
\begin {equation} 
        \eta^\mu 
        = (\eta^+,\vec\eta_{\!\perp},\eta^-)
        = (0,\vec 0_{\!\perp},2)
\ ,     \label{eq:e100}\end {equation}
with $\eta^2= 0$. The identities
\begin{eqnarray}      
      {\vec k _{g_{\!\perp}} ^{\,2} \over x_g} 
      &=& (k_g+k_q+k_{\bar q}^\prime)^2 - (k_q+k_{\bar q})^2 
      = - {1\over x_g}\left(k_{\bar q}^\prime - k_{\bar q} \right)^2
\\   &=& (k_g+k_q+k_{\bar q}^\prime)^2 -
      (k_q^\prime+k_{\bar q}^\prime)^2
      = - {1\over x_g}\left(k_q-k_q^\prime\right)^2
\   \end{eqnarray}
are then obtained straightforwardly. 

Some words of caution seem to be in place. 
(1) The smallness condition requires that $\widetilde\Lambda$ 
is smaller than the lowest bound state.
Restraining the transverse momentum of the gluon to be
less than the pion mass is not a serious draw-back. One
should  note that the off-shell mass of the gluon can be large 
irrespective of how well the smallness condition is fulfilled. 
(2) One can state with certainty, that the above reduction is
inadequate if not false at sufficiently large values of $\omega$, 
where one might be in the  region of `overlapping resonances'  
\cite{vwz85}. This will require different technologies
and possibly   those based on random matrix models 
\cite{ver94,wsw96}. 
It is there where the concept of a {\em temperature} will find its 
way into the bound state problem of gauge field theory.

Finally, the resolvent in the $q\bar q\,g$-sector has
determined itself to be very simple:
{\em In the solution, the resovents $\overline G _n$ are 
diagonal in Fock-space representation and strictly 
independent of $\omega$.
Instead of having `resolvents of resolvents' the hierarchy
of resolvents is broken.
The particles `in medium' propagate like free particles.}
\section{The effective interaction for Gauge Theory}
\label{sec:7}
\begin{figure} [t]
\begin{minipage}[t]{78mm} 
\makebox[0mm]{}
\epsfxsize=78mm\epsfbox{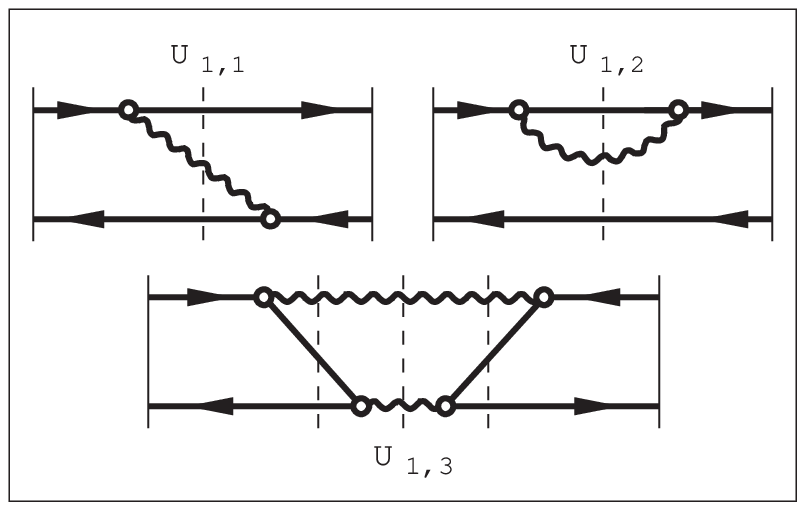}
\caption{\label{fig:7.1} \sl
   The full effective interaction in the $q\bar q$-space,
   resumed to all orders in the bare coupling constant.
   Circles represent the running coupling constant.
   The one-gluon exchange interaction 
   $U\equiv U_{1,1} + U_{1,2}$ provides the binding.
   The annihilation interaction $U_a\equiv U_{1,3} $ can be
    active only     if the quarks have the same flavor. 
} \vfill \end{minipage}
\hfill
\begin{minipage}[t]{78mm} 
\makebox[0mm]{}
\epsfxsize=78mm\epsfbox{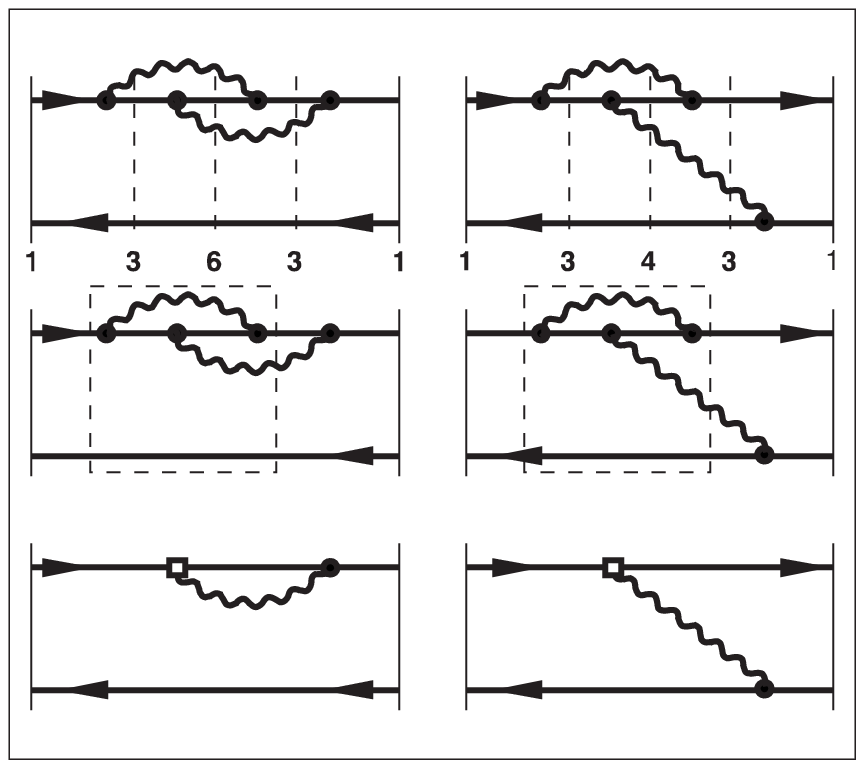}
\caption{\label{fig:7.2} \sl
   Two typical graphs, taken from Figure~\protect\ref{fig:6_4}, 
   should illustrate how the propagator box (dashed line)  
   contracts into a square around the vertex.
} \vfill \end{minipage}
\end{figure}

The above considerations are summarized as follows. 
The effective Hamiltonian 
\begin  {equation} 
       H _{\rm eff}= T + U + U_a
\   \label {eq:7.0.01} \end {equation} 
is supposed to be diagonalized {\em only in the
$q\bar q$-space}.   The eigenvalue equation 
$ H_{\rm eff} \,\vert \psi_b \rangle = E_b \,\vert \psi_b \rangle $ 
can be understood as the finite dimensional ($N_s$)
matrix equation
\begin {equation} 
       \sum _{j=1} ^{N_s} \langle i \vert H _{\rm eff} \vert j \rangle 
       \langle j \vert\psi_b \rangle 
       =  E_b  \ \langle i \vert\psi_b \rangle 
\ , \ \quad{\rm for}\ i=1,2,\dots,N_s
. \label {eq:7.0.02} \end {equation}
Subject to fixed   values of $P ^+$ and  $\vec P _\perp $,  
the basis states $ \vert j \rangle $ denumerate all possible
Fock states   
\begin {equation} 
        \vert j \rangle =  \vert q, \bar q \rangle 
       = b^\dagger_{q} d^\dagger _{ \bar q} \vert 0 \rangle
       ,\qquad{\rm or}\quad \vert q, \bar q \rangle
       = \vert x,\vec k_{\!\perp}; \lambda_{q}
        , \lambda_{\bar q} \rangle 
\ .    \label {eq:7.0.03} \end {equation}
It often suffices to label an individual
Fock state by the momenta of the quark $x\equiv x_{q}$ and 
$\vec k_{\!\perp} \equiv \vec k_ {q_{\!\perp}}$, 
since the antiquark has
$x_{\bar q} = 1-x$ and 
$\vec k_ {\bar q_{\!\perp}} = -\vec k_{\!\perp} $. 
The effective Hamiltonian has a kinetic energy $T$ and two
kinds of potential energies $U$. As operators acting in Fock 
space they are defined by
\begin {eqnarray}
       T &=&    \sum_{q} b^\dagger_{q} b _{q} 
       \left({ m ^2+\vec k _{\!\bot}^{\,2}\over x} \right) _q
      +  \sum_{\bar q} d^\dagger_{\bar q} d _{\bar q} 
       \left({ m ^2+\vec k _{\!\bot}^{\,2}\over x} \right) _{\bar q}
\ ,   \label {eq:7.0.04} \\  
      U &=& \sum_ {q, \bar q ,  q ^\prime , \bar q^\prime } 
      \ b^\dagger_{q} d^\dagger _{\bar  q }  
        d_{\bar  q^\prime } b_{q^\prime } 
      \ \langle q, \bar q \vert 
      \, \widetilde U \, \vert q^\prime , \bar q^\prime \rangle 
      \ \delta _{f_q,f_{q^\prime}} 
      \ \delta _{f_{\bar  q},f_{\bar  q^\prime}} 
\ ,  \label {eq:7.0.05}\\  
       U _a &=& \sum_ {q, \bar q ,  q ^\prime , \bar q^\prime } 
      \ b^\dagger_{q} d^\dagger _{\bar  q }  
        d_{\bar  q^\prime } b_{q^\prime } 
      \ \langle q, \bar q \vert 
      \, \widetilde U _a\, \vert q^\prime , \bar q^\prime \rangle 
      \ \delta _{f_q,f_{\bar  q}} 
      \ \delta _{f_{q^\prime},f_{\bar  q^\prime}} 
\ .   \label {eq:7.0.06}\end {eqnarray} 
One distinguishes a flavor-conserving ($U$) and a 
flavor-changing effective interaction ($U_a$). 
Both of them scatter a quark from state $q$ to state
$q^\prime$,  and the antiquark from $\bar q$ to $\bar q ^\prime$.
They  are given diagrammatically in Figure~\ref{fig:7.1}, and
more formally by 
\begin  {eqnarray} 
       U      &=& V \,G _3\,  V
                   =    V \,R_3^\dagger  \overline G _3  R_3\, V
\ , \qquad\qquad{\rm and }\label {eq:7.0.07}  \\
       U_a &=& V \,G _3\,  V G _2  V  \,G _3 \,V 
                   =    V \,R_3^\dagger \overline G _3 R_3\,  V  
       G _2  V \,R_3^\dagger  \overline G _3 R_3\,V 
\ . \label {eq:7.0.08} \end {eqnarray}
Most importantly, the propagator $\overline G _3 $
(but {\em not} $G _3 $!) `in the solution' is {\em diagonal in 
Fock space representation}, its numerical value being
closely related to the momentum transfer $Q$ of the quarks, 
\begin  {equation} 
       \langle q;\bar q; g \vert \overline G _3  \vert q^{\prime\prime};
       \bar q^{\prime\prime}; g^{\prime\prime} \rangle 
  =   {x_g\over Q^2}
       \ \langle q;\bar q; g \vert q^{\prime\prime};
       \bar q^{\prime\prime}; g^{\prime\prime} \rangle
       \ ,\qquad\qquad{\rm with}\quad
      Q^2 = (k_q - k_q^\prime)^2
\ . \label {eq:7.0.09} \end {equation} 
The operators $R$ had been defined in Eq.(\ref{eq:6510});
here they read 
\begin  {equation}
       R  _3  =  {1\over \sqrt { 1 - \overline G _3 \widetilde U _3}  }
\ .  \label {eq:7.0.10} \end {equation} 
What is their interpretation?

\subsection{The Running Coupling Constant}
\label{sec:7.1}

The role of  $R$ becomes more obvious when setting 
$R=1$ but keeping $R^\dagger $ in Eq.(\ref{eq:7.0.07}), 
\begin  {equation}
       U   \simeq   V R_3^\dagger  \overline G _3  V 
\ .  \label {eq:7.1.01} \end {equation} 
Expanding  $R^\dagger $ to first non-trivial order,
picking out a particular term,  one gets diagrams like in
Figure~\ref{fig:7.2}, or formally 
\begin  {equation}
       U   \simeq   V \overline G _3   V + 
             {1\over 2} V  \overline G _3   V \overline G _6 V
\overline G _3 V + \dots\ .
\   \label {eq:7.1.02} \end {equation} 
As mentioned already, a propagator should be
represented rather by a box than by a vertical line.
This becomes important when the propagator is diagonal,
see Eq.(\ref{eq:7.0.09}), {\it i.e.} when it does not change 
the momenta of the  `in-' and `out-states'. 
As shown in Figure~\ref{fig:7.2}, the surrounding box lines 
can then be contracted over the exchanged gluon. 
The in- and out-momenta of the $R$-box are then determined
only by the quark single-particle four-momenta $k_q$  and 
$k_q^\prime$. Note that these two also determine uniquely 
the four-momentum of the irradiated gluon. 
Therefore, in Fock-space representation, the operator 
$R$ must be a {\em function of $k_q$  and $k_q^\prime$}
alone, and thus only a function of the four-momentum 
transfer $Q$ as defined in Eq.(\ref{eq:7.0.09}). 
The {\em vertex function}
\begin  {equation}
       \langle k_q \vert R \vert k_q ^\prime, k_g \rangle 
       = r (k_q,k_q^\prime;\Lambda, m_q, g)
       = r (Q,\Lambda)  
\   \label {eq:7.1.05} \end {equation} 
depends thus on all variables in the problem, particularly on 
$Q$ and the cut-offs $\Lambda$.  
As the figure demonstrates,  the operator $R$ and its matrix
elements {\em are closely related to the running 
coupling constant}: both have the same perturbative 
expansion \cite{grw73,pol73}. But it should be emphasized that perturbative
expansions are used here only for the purpose of illustration.
The present formalism, particularly Eq.(\ref{eq:7.0.10}),
resumes it to all orders in the bare coupling  constant, 
contrary to the familiar expressions 
gained from perturbation theory \cite{grw73,pol73}.
One should mention here the potential danger hidden in
the cut-off depence of $r (Q,\Lambda)$. Perturbative
estimates yield diverging expressions for ever increasing
value of  $\Lambda$. The problems related to these
well-known divergencies will be discussed in future work,
see also section~\ref{sec:7.4}.

\subsection{The effective interaction}
\label{sec:7.2}

The actual computation of the vertex function   
$r (Q,\Lambda)$ as function of all its 
parameters clearly goes beyond the scope of the present
work and must be left to the future, despite its importance.
By the same reason, the computation of the annihilation
interaction can not be given here. Rather shall we restrict
ourselves here to calculate explicitly the flavor-conserving 
interaction $U$.

Consider the matrix element  $U_{1,1}$ in 
Figure~\ref{fig:7.1}. In this graph, an initial state 
$\vert q ,\bar q \rangle $ state is scattered into a final
state $\vert q^\prime, \bar q^\prime \rangle $,
going through an intermeditate state 
$\vert q^\prime, \bar q,g  \rangle $.
As an adventage of DLCQ, one can introduce these
states as to be ortho-normalized and invariant under 
$SU(N)$: 
\begin  {eqnarray}
       \vert q ,\bar q \rangle 
       &=&{1\over \sqrt{n_c}} \sum _{c=1} ^{n_c}
       b^\dagger_c (k_q,\lambda_q) 
       d^\dagger_c  (k_{\bar q},\lambda_{\bar q})
       \vert vac \rangle 
\  ,  \label {eq:7.2.01} \\ 
       \vert q^\prime, \bar q^\prime \rangle 
       &=&{1\over \sqrt{n_c}} \sum _{c=1} ^{n_c}
       b^\dagger_c (k_q^\prime,\lambda_q^\prime) 
       d^\dagger_c  (k_{\bar q}^\prime,\lambda_{\bar q}^\prime)
       \vert vac \rangle \  ,\qquad\qquad{\rm and}
\     \label {eq:7.2.02} \\ 
      \vert q^\prime, \bar q,g  \rangle 
&=& \sqrt{{2\over n_c^2-1}} 
       \sum _{c=1} ^{n_c} 
       \sum _{c^\prime=1} ^{n_c} 
       \sum _{a=1} ^{n_c^2-1} 
       T^a_{c,c^\prime}
       b^\dagger_c (k_q^\prime,\lambda_q^\prime) 
       d^\dagger_c  (k_{\bar q},\lambda_{\bar q})
       a^\dagger_a  (k_{g},\lambda_{g})
       \vert vac \rangle 
\  .  \label {eq:7.2.03} \end {eqnarray} 
The color matrices $T^a_{c,c^\prime}$ are normalized as 
usual by $2{\rm Tr} (T^a T^b) = \delta _{a,b}$.
Since the gluon has 
a positive value of $k_q^+$, one necessarily has either 
\begin  {equation}
      k_q^+ > k_q^{\prime+} 
      \qquad{\rm or}\qquad\qquad
      k_{\bar q}^+ > k_{\bar q}^{\prime+} 
\  .  \label {eq:7.2.04} \end {equation} 
We restrict here to the former. 
The calculation of the vertex interaction proceeds in two steps. 
By means of the tables \cite{brp91},  one calculates the vertex 
matrix elements, separately 
for emission and absorbtion of the gluon, 
\begin {eqnarray} 
       \langle q ,\bar q \vert \,VR^\dagger\,
       \vert q^\prime, \bar q , g  \rangle 
       &=& \sqrt{{n_c^2-1\over 2n_c}}
        \ \sqrt{{g^2 r^2(Q,\Lambda) \over \Omega\,P^+}}
        \ {\left[ \overline u (k_q,\lambda_q) 
        \,\gamma^\mu\epsilon_\mu (k_g,\lambda_g)\,
        u(k_q^\prime,\lambda_q^\prime)\right] 
        \over \sqrt{x_{q}x_{q}^\prime x_{ g}}}
\  ,  \label {eq:7.2.06} \\
       \langle q^\prime, \bar q ,g  \vert \,RV\,
       \vert q ^\prime ,\bar q ^\prime \rangle 
       &=& -\sqrt{{n_c^2-1\over 2n_c}}
        \ \sqrt{{g^2 r^2(Q,\Lambda) \over \Omega\,P^+}}
        \ {\left[ \overline u (k_{\bar q},\lambda_{\bar q}) 
        \,\gamma^\nu\epsilon^\star _\nu(k_g,\lambda_g)\,
        u(k_{\bar q}^\prime,\lambda_{\bar q}^\prime)\right] 
        \over \sqrt{x_{\bar q} x_{\bar q}^\prime x_{ g}}}
\  ,  \label {eq:7.2.07} \end {eqnarray} 
respectively. 
Since both the (LC)-Hamiltonian and the Fock states are color
neutral, all color algebra has reduced into the overall factor
$(n_c^2-1)/2n_c$. 
According to Eq.(\ref{eq:7.0.07})
one has to sum over all intermediate states. 
Since the momenta are fixed, only the sum over the two 
gluon helicities remains, which is done in the usual way:
\begin {eqnarray} 
        d_{\mu\nu}(k_g) \equiv \sum_{\lambda_g}
        \epsilon_\mu (k_g,\lambda_g)\,
        \epsilon^\star_\nu (k_g,\lambda_g)
        = -g_{\mu\nu} + 
        {k_{g,\mu}\eta_\nu + k_{g,\nu}\eta_\mu 
        \over k_g^\kappa \eta_\kappa } 
\ .    \label {eq:7.2.08} \end {eqnarray} 
The time-like vector $\eta$ had been defined in 
Eq.(\ref{eq:e100}).
In Appendix~\ref{sec:g} it is shown how the gauge remnant
of the polarization sum cancels exactly against the
contribution from the instantaneous interaction. 
The latter, remember, appears since one works  
in the light-cone gauge $A^+=0$.
The remainder due to $g_{\mu\nu}$ becomes then 
{manifestly gauge invariant}:
\begin{eqnarray} 
        \langle q, \bar q \vert 
        \widetilde U \vert q^\prime, \bar q ^\prime \rangle 
        &=&  {2(2\pi)^3\over  \Omega P ^+} 
        \ \langle q, \bar q \vert 
        U \vert q^\prime, \bar q ^\prime \rangle 
\ ,     \label{eq:7.2.21} \\ {\rm with}\quad  
        \langle q, \bar q \vert 
        U \vert q^\prime, \bar q ^\prime \rangle 
        &=& -{1\over 4\pi^2}\ {\beta(Q,\Lambda) \over Q  ^2} 
        \ {\langle \lambda_q,\lambda_{\bar q} 
        \vert S(Q) \vert 
        \lambda_q^\prime,\lambda_{\bar q}^\prime \rangle 
        \over \sqrt{x _q(1 - x  _q)\,x _q^\prime (1- x _q^\prime)}}
\ ,    \label {eq:7.2.22} \\ {\rm and}\quad 
        \langle \lambda_q,\lambda_{\bar q} \vert S(Q) \vert 
        \lambda_q^\prime,\lambda_{\bar q}^\prime \rangle 
        &=& \left[ \overline u (k_{q},\lambda_{q}) 
        \,\gamma^\mu\,
        u(k_{q}^\prime,\lambda_{q}^\prime)\right] 
        \left[ \overline u (k_{\bar q},\lambda_{\bar q}) 
        \,\gamma_\mu\,
        u(k_{\bar q}^\prime,\lambda_{\bar q}^\prime)\right] 
.      \label {eq:7.2.23} \end{eqnarray} 
The spinor factor $S(Q)$ collects the familiar and 
{\em manifestly Lorentz invariant} current-current 
interaction, which ultimately will be responsible for the 
fine and hyperfine structure.

The bare coupling constant $g$  is combined with the
vertex function $r(Q,\Lambda)$ into the like-to-be 
`running coupling constant' $ \beta(Q,\Lambda) $, 
\begin {equation} 
        \beta(Q,\Lambda) =  {n_c^2-1\over 2n_c}
        \  {g^2\over 4\pi \hbar c}  \  r^2(Q,\Lambda)
\ . \label {eq:7.2.24} \end {equation} 
For QCD the gauge group factor $(n_c^2-1)/2n_c$
reduces to $4/3$, for QED it has the value $1$.
It is here where the essential difference of Abelian and 
non-Abelian gauge theory appears: The coupling
constant runs differently. As a consequence, the one is
confining and the other one is not.~---  The graph 
$U_{1,2}$ in Figure~\ref{fig:7.1} is computed
correspondingly. 

In the discretized case of the matrix equation, no
divergencies can occur because  of the regularization. 
In practical work, however, and by matter of principle, it is
more convenient to go to the {\em continuum limit} 
$K\rightarrow\infty$, simply by
replacing sums with integrals. 
The conversion factor from Eq.(\ref{eq:316}) cancels 
against the corresponding factor in Eq.(\ref{eq:7.2.21}), 
and the {\em matrix equation} (\ref{eq:7.0.02}) is 
converted into the {\em integral equation} 
\begin {eqnarray} 
        M_b^2 
        \langle x,\vec k_{\!\perp}; \lambda_{q},
        \lambda_{\bar q}  \vert \psi _b\rangle 
        = \left[ {m_{q} + \vec k_{\!\perp}^2 \over x } 
        + {m_{\bar q} + \vec k_{\!\perp}^2 \over 1-x } \right]
        \langle x,\vec k_{\!\perp}; \lambda_{q},
        \lambda_{\bar q}  \vert \psi _b\rangle 
\nonumber\\
        + {1\over 4\pi^2}
        \sum _{ \lambda_q^\prime,\lambda_{\bar q}^\prime }
        \!\!\! \int^{\Lambda}
        \!\!\! dx^\prime d^2 \vec k_{\!\perp}^\prime
        \,{\beta(Q,\Lambda) \over Q  ^2} 
        \,{\langle \lambda_q,\lambda_{\bar q} 
        \vert S(Q) \vert 
        \lambda_q^\prime,\lambda_{\bar q}^\prime \rangle 
        \over \sqrt{x _q (1 - x  _q) x _q^\prime (1- x _q^\prime)} }
        \left[ \langle x,\vec k_{\!\perp}; 
        \lambda_{q},\lambda_{\bar q}  
        \vert \psi _b\rangle 
        - \langle x^\prime,\vec k_{\!\perp}^\prime; 
        \lambda_{q}^\prime,\lambda_{\bar q}^\prime  
        \vert \psi _b\rangle \right]
.      \label {eq:7.2.26} \end {eqnarray}
The $\Lambda$ should remind to the fact that the 
integration over the perpendicular momenta is restricted 
to a range consistent with Eqs.(\ref{eq:220}) and (\ref{eq:e80}). 
Up to the omitted annihilation term, the resulting eigenvalues 
$M_b^2$ {\em must} coincide with the eigenvalues of the full 
Hamiltonian as shown above. The mass 
$M_b$ {\em must be interpreted} as the mass of a 
physical meson. The corresponding wavefunctions 
$      \langle x,\vec k_{\!\perp}; \lambda_{q},
        \lambda_{\bar q}  \vert \psi _b\rangle$ 
give the probability amplitudes for finding in that meson 
a quark with momentum (fraction) $\vec k_{\!\perp}\,(x)$ and
helicity $\lambda_{q}$. 
Finally, one should emphasize that the front-form 
effective Hamiltonian, appearing as the  
kernel of the integral equation (\ref{eq:7.2.26}), 
is {\em manifestly boost invariant}. 

\subsection{Renormalization Group analysis}
\label{sec:7.4}

The integral equation (\ref{eq:7.2.26}) 
can be solved for any fixed value of the cut-offs
$\Lambda$.
Therefore, both the wavefunctions and the eigenvalues 
depend explicitly on it, $M_b = M_b(\Lambda)$.

This is unphysical, since a physical results should not
depend on mathematical tricks like regularization.
Beyond that it is potentially dangerous since the vertex
function diverges as a function of $\Lambda$.
The way out is, a  future renormalization group analysis. 
One has to require that the solutions do not depend on
the cut-off
\begin {equation} 
        d\,M_b(\Lambda;m, \beta;Q) = 0,
        \qquad{\rm or}\quad
        \delta_{\Lambda}M_b\,+\delta_{m}M_b\,
        +\delta_{\beta}M_b=0 
\ , \label {eq:7.4.01}\end {equation}
at a particular scale set by the momentum transfer $Q^2$.
The renormalization group analysis has thus far been done
only for asymptotically large 
$Q^2\rightarrow\infty$. It is a well established fact, even
in light-cone quantization 
\cite{leb81,brp91,tho79,phz93},  that asymptotically holds
$ \beta(Q) \longrightarrow \beta_0 /\ln(Q ^2/\kappa ^2)$, 
with a coefficient $\beta_0$ well-known from theory and a
mass scale $\kappa$ which must be  determined from 
experiment.
The only difference between  the textbooks and the present 
approach is, that the  coupling function $\beta(Q)$ must be
calculable down to momentum-transfer zero. The latter
is required when one solves the integral equation. 
Here is the problem, and here we must leave it, 
since the explicit calculation of $r(Q,\Lambda)$ 
and the subsequent renormalization of 
$\beta(Q,\Lambda)$ to yield a $\beta(Q,\kappa)$
clearly goes beyond the scope of the present work.
But we have spotted where the problem is.

\subsection{Retrieving the full wavefunction}
\label{sec:7.3}

One should emphasize finally that
the knowledge of the $q\bar q$-space eigenfunction
$\psi_b$ is  sufficient to retrieve all desired Fock-space  
components of the total wavefunction $\Psi$. 
The key is the upwards recursion relation
Eq.(\ref{eq:411}),  {\it i.e.}
\begin {equation} 
        \langle n \vert \Psi \rangle   =
        \sum _{j=1} ^{n-1} G _ n 
        \langle n \vert H _n\vert j \rangle 
        \langle j \vert \Psi  \rangle 
\ . \label {eq:7.3.a}\end {equation}
Obviously, one can express the higher Fock-space components 
$\langle n \vert \Psi \rangle$ as functionals of $\psi$ 
by a finite series of quadratures (or matrix multiplications). 
One does not have to solve another eigenvalue problem.  
In order to show this, we ask for the probabilty amplitude 
to find a  $gg(2)$ or a  $q\bar q\,g(3)$ state in a particular 
meson $b$ as an example.
With
$\langle 1 \vert \Psi \rangle 
 =  \langle q\bar q \vert \psi_b\rangle$ 
the first two relevant equations of (\ref{eq:7.3.a}) are
\begin {eqnarray} 
        \langle 2 \vert \Psi \rangle  &=& 
        G _ 2 \langle 2 \vert H _2 \vert 1 \rangle 
        \langle 1 \vert \Psi  \rangle 
\ , \qquad\quad{\rm and } \label {eq:7.3.03} \\  
        \langle 3 \vert \Psi \rangle  &=& 
        G _ 3 \langle 3 \vert H _3 \vert 1 \rangle 
        \langle 1 \vert \Psi  \rangle +
        G _ 3 \langle 3 \vert H _3 \vert 2 \rangle 
        \langle 2 \vert \Psi  \rangle 
\ . \label {eq:7.3.04}\end {eqnarray}
Note that $\langle 2 \vert \Psi \rangle$ is expressed already 
in terms of $\langle 1 \vert \Psi \rangle$. 
The sector Hamiltonians $H _n$ 
have been dealt with in sections~\ref{sec:4} and~\ref{sec:6}.
They can be expressed by chains with the bare interaction
$V$,  particularly  
$H _2 = T + V + VG _3V + VG _5V$ and 
$H _3 = T + V + VG _4V + VG _6V + VG _6VG _5V G _6V$.
With the abbreviation 
$V_{ij}\equiv\langle i\vert V\vert j\rangle$ and the 
`selection rules' of Figure~\ref{fig:holy-2} one gets 
$\langle 2\vert H _2 \vert 1 \rangle =  V_{23} G _3 V_{21}$,
$\langle 3\vert H _3 \vert 1 \rangle =  V_{31} $, 
$\langle 3\vert H _3 \vert 2 \rangle =  V_{32} $, and therefore
\begin {eqnarray} 
        \langle 2 \vert \Psi \rangle  &=& 
        G _ 2 V_{23} G _3 V_{21} \langle 1 \vert \Psi  \rangle  , 
\qquad\quad{\rm and } \label {eq:7.3.07} \\  
        \langle 3 \vert \Psi \rangle  &=& 
        G _ 3 V_{31} \langle 1 \vert \Psi  \rangle 
    + G _ 3 V_{32} G _ 2 V_{23} G _3 V_{21} 
        \langle 1 \vert \Psi  \rangle 
\ . \label {eq:7.3.08}\end {eqnarray}
One should emphasize again that these expressions are
exact to all orders in the coupling constant.
The propagators have to be treated like exposed in 
section~\ref{sec:6}. They include the running coupling
constant and in the solution are independent of $\omega$.
Obviously, the procedure is straightforward, well-defined,  
and actually simple.

\section {Summary and Perspectives} 
\label {sec:8}

In the present work, one adopts the point of view that 
most, if not all, properties of a Lagrangian gauge field 
theory are contained in the canonical front-form 
Hamiltonian, which is calculated in the  light-cone gauge 
omitting the zero modes.
Periodic boundary conditions allow to
construct explicitly the Hamiltonian matrix for fixed
harmonic resolution.   
Rows and columns of this matrix refer to all
possible momentum states of  2, 3, or more  
particles. All variables are well defined, finite and 
denumerable: The total longitudinal momentum and  
the harmonic resolution are finite, the number of states 
and  the number of sectors are finite, and last but not least 
every single matrix element of the Hamiltonian is  finite. 
One is confronted with the diagonalization of a finite 
block matrix and applies the theory of effective interactions 
in conjunction with the method of iterated resolvents.
As shown, one can map the large Hamiltonian matrix  
without smallness assumption onto  a matrix 
problem with much smaller dimensions, to the problem of 
diagonalizing the effective  Hamiltonian in the  $q\bar
q$-space.  It is shown why {\em these  two matrices  
have the same eigenvalues}. 
At the end, the periodic boundary conditions are relaxed 
by the limiting process of going over to the continuum limit.

The diagrammatic representation of the effective interaction
as given in Figure~\ref{fig:7.1}  looks like second order 
diagrams of perturbation theory. But despite this, they are far
from  being that: due to the 
vertex function (which after renormalization is related to the
running coupling constant), they represent a resummation of
pertubative graphs to all orders in the bare coupling constant. 
The effective interaction $U$ is computed 
explicitly in section~\ref{sec:7}.
It is our understanding that $U+U_a$ represents an exact
mapping of the full many-body Hamiltonian onto the 
$q\bar q$ space. Solving for the eigenfunctions in this space
one can retrieve all other many-body amplitudes of the
wavefunction in a self-consistent  way, as shown 
in section~\ref{sec:7}.

To arrive at this simple result it is crucial to
include arbitrarily many gluons. Only then, as argued in 
section~\ref{sec:6}, the Russian puppet structure of 
`resolvents within resolvents' can be used as an argument
why the effective interaction is the same in all hierarchies.
It is precisely this aspect of self-similarity in a gauge theory
which ultimately allows for the `breaking of the hierarchy':
In the solution, the particles propagate like free particles and
all many-body aspects reside in the vertex coupling function.
One can overstress the point for QED by stating that the
generation of the simple  Coulomb potential requires 
infinitely many photons. Finally, one should emphasize that
the present approach needs no `convergence-improving'
external potential as in \cite{wwh94}, which once introduced is so
difficult to get rid-off. 
The average potential generates itself, in the solution.

The present work claims to connect a Lagrangian
gauge field theory with the eigenvalues and eigenfunctions 
of a bound state equation. 
We like to emphasize that all steps can be 
(dis-)verified numerically in a well defined manner. One can
start with a sufficiently large DLCQ-Hamiltonian matrix, 
as large as the computer can digest. 
The reduction to the effective Hamiltonian 
can then be done by successive matrix inversions and
multiplications as described in Sections~\ref{sec:3}
and \ref{sec:4}.
The prediction is that the lower eigenvalues depend only
weakly on the starting point energy. In addition, one can
(dis-)verify  numerically {\it a posteriori} how well the
succesive approximations  from Eqs. (\ref{eq:57}) to 
(\ref{eq:58}) to (\ref{eq:59}) are satisfied in practice. 

Important as the present work might be, it only can be 
the first step. The resulting effective potential depends
on several formal and unphysical parameters, among them the
harmonic resolution $K$ and the transversal length 
$L_\perp$ induced both by the periodic boundary conditions,
the cut-off scale $\Lambda$ regulating the transversal
momenta, and finally the smallness parameters $\epsilon$
and $\widetilde\Lambda$.
The resolution and the length disappear when going to
the continuum limit. But the dependence on $\Lambda$,
$\epsilon$ and $\widetilde\Lambda$ remains. The second, 
and perhaps even more  important step, must therefore be 
to remove them by a renormalization group analysis 
\cite{grw73,pol73}. This has not been done, yet.
The present work thus culminates and ends at a
pre-renormalization stage. 
The merit of the present work is to pinpoint 
the object of renormalization, namely the vertex function
which is given here for the first time in a non-perturbative 
formulation to all orders in the coupling constant.

But even before a future renormalization group analysis the
present formalism is useful and has some rather nice aspects:
Even without knowing the explicit structure of the running
coupling constant in the infinite momentum frame one can do
some educated guess work. For example, using
Richardson's version \cite{ric79} of the running coupling 
constant one can establish confinement in a parameter-free 
fashion. One can approximately, but analytically, calculate
the heavy meson masses based on the
parametric variation of a trial wave function \cite{pam95}. 
An explicit, but numerical, solution of the same problem is
currently being attempted \cite{trp96}. Last but not least one
can choose the smallness parameters $\widetilde\Lambda$ 
so small that no transversal momenta survive at all. 
One then ends up with a `collinear model' \cite{dkp94} 
similar to the dimensionally reduced
models solved thus far~\cite{pkp95,pab96,vab96},
resulting in stupendous
analytical properties of the light mesons \cite{bap96}.
These features make it worth to communicate
the present work even prior to a
renormalizaton group analysis.

\underline {Acknowledgement}. 
I am grateful to Uwe Trittmann and Rolf Bayer, who 
have devoted much time to help me in the early phases of 
this work. In particular, I thank Uwe for preparing 
Figure~\ref{fig:4_1} for me, among many others not shown. 
I thank my friends Profs. Steve Pinsky from OSU and 
Stan Brodsky from SLAC for their patience and help to
improve my never ending illiteracy on field theory off and on
the light cone.

\begin{appendix}
\section {The $4\times 4$ Block Matrix as a Paradigm}
\label {sec:b}

The considerations in section~\ref{sec:4} can be substantiated 
with an explicit example. The rows and columns of any finite
dimensional matrix can be grouped into four blocks.
The block matrix has then the following shape: 
\begin {equation}  
  \pmatrix{ \langle 1 \vert H \vert 1 \rangle 
          & \langle 1 \vert H \vert 2 \rangle 
          & \langle 1 \vert H \vert 3 \rangle 
          & \langle 1 \vert H \vert 4 \rangle 
\cr         \langle 2 \vert H \vert 1 \rangle 
          & \langle 2 \vert H \vert 2 \rangle 
          & \langle 2 \vert H \vert 3 \rangle 
          & \langle 2 \vert H \vert 4 \rangle 
\cr         \langle 3 \vert H \vert 1 \rangle 
          & \langle 3 \vert H \vert 2 \rangle 
          & \langle 3 \vert H \vert 3 \rangle 
          & \langle 3 \vert H \vert 4 \rangle 
\cr         \langle 4 \vert H \vert 1 \rangle 
          & \langle 4 \vert H \vert 2 \rangle 
          & \langle 4 \vert H \vert 3 \rangle 
          & \langle 4 \vert H \vert 4 \rangle\cr} 
 \ . \label {eq:501b} \end {equation}
The reduction from the $4\times 4$ to the $3\times 3$ matrix is easy:  
Since $ H_4 \equiv H$, one replaces all bare block matrix elements 
in the $3\times 3$ matrix 
by $ \langle i \vert H _3\vert j \rangle $,  with  
\begin {equation}  
    H _3 =   H  +  H G _4 H 
\ .\label {eq:502b} \end {equation} 
Next, reduce the matrix of block matrix 
dimension 3 to the one of block matrix dimension 2, by
$  H _2 =   H_3  +  H _3  G _3 H _3 $.
Inserting $ H _3 $ from Eq.(\ref{eq:502b}), one gets 
\begin {equation}  
    H _2 = ( H  +  H G _4 H ) 
         + ( H  +  H G _4 H ) G _3 ( H  +  H G _4 H ) 
\ . \label {eq:504b} \end {equation} 
Performing reduction and substitution once more one arrives at
\begin {eqnarray} 
   H _1 
   & = & 
        \Bigl( ( H  +  H G _4 H ) 
      + ( H  +  H G _4 H ) G _3 ( H  +  H G _4 H ) \Bigr) G _2 
 \nonumber \\ & \phantom{=} & 
        \Bigl( ( H  +  H G _4 H ) 
      + ( H  +  H G _4 H ) G _3 ( H  +  H G _4 H ) \Bigr) 
\ . \label {eq:506b} \end{eqnarray} 
One has thus expressed the matrix elements of the 
effective interaction in sector 1 in terms of the bare interaction
$ H $ and the resolvents $G _2 $, $ G _3 $, and $ G _4 $, which in turn 
are given by the effective interactions  $ H _2 $, $ H _3 $ and 
$ H _4 $, respectively, {\it i.e.}
\begin {equation} 
G _2 = \vert 2\rangle\,{1\over \omega-H _2}\,\langle 2 \vert 
\ , \quad
G _3 = \vert 3\rangle\,{1\over \omega-H _3}\,\langle 3\vert
\ , \quad
G _4 = \vert 4\rangle\,{1\over \omega-H}\,\langle 4\vert
\ . \label {eq:507b} \end {equation} 
Note that all operations are well defined matrix multiplications 
and  inversions. The {\em bare interactions} $H$ alternate
with the resolvents  of {\em effective interactions} $ H_n$ to
build up {\em strictly finite chains}. The longest chain in 
Eq.(\ref{eq:506b}) has  5 propagators: 
$ H G _4 H G _3 H G _2 H G _3 H G _4 H $. 
This is in stark contrast  to the infinite chains of perturbative
series.  

\section{ Continued Fractions in Iterated Resolvents} 
\label {sec:c}
\begin{figure} [t]
\begin{minipage}[t]{80mm} 
\makebox[0mm]{}
\epsfysize=80mm\epsfbox{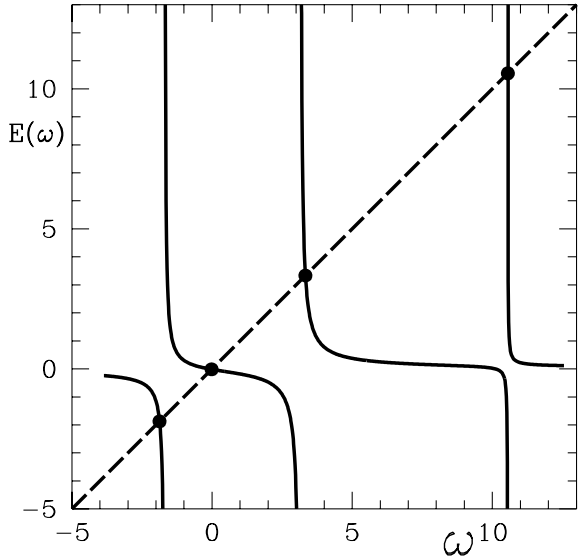}
\makebox[0mm]{}
\end{minipage}
\hfill
\begin{minipage}[t]{80mm} 
\makebox[0mm]{}
\caption{\label{fig:4_1} \sl 
    The energy function  $E(\omega)$  for the 
    matrix $H_B$ is plotted versus $\omega$.
    The solutions of  $ E(\omega) =  \omega  $ agree with 
    the four exact eigenvalues 
    -1.87208, -0,01518, 3.33343, and 10.5538.
} \vfill \end{minipage}
\end{figure}
The structure of the chains in the Method of Iterated
Resolvents depends quite strongly on the block matrix
structure of the matrices considered, as is obvious from
Eq.(\ref{eq:506b}) in Appendix~\ref{sec:b}. For to show that
explicitly, the example of a matrix with a tridiagonal block
structure is choosen, {\it i.e.}
\begin {equation}  
  H _B =
  \pmatrix{ H_{11}           & H_{12}           & .          & .
\cr         H_{21}           & H_{22}           & H_{23}          & . 
\cr         .          & H_{32}           & H_{33}           & H_{34} 
\cr         .           & .           & H_{43}           & H_{44} \cr} 
\ , \qquad {\bf or} \quad 
  \widetilde H _B =
          \pmatrix{ 0 & 1 & . & . \cr 
            1 & 2 & 3 & . \cr
            . & 3 & 4 & 5 \cr 
            . & . & 5 & 6 \cr } 
\ , \label {eq:508c} \end {equation} 
with dots representing zero matrices ($H _B$) or straight 
zeros ($\widetilde H _B $).
Applying the same procedure as in Appendix~\ref{sec:b}, one
notes that many of the long chains vanish since for example
$G _2 H G _4 \equiv  
   G _2 \, H _{24}\, G _4 = 0 $. 
The effective interactions, as given in Eq.(\ref{eq:506b}) 
for the general $4\times4$ block matrix, simplify strongly. 
Here, one gets consecutively 
$    H _3 = H + H G _4 H $,
$    H _2 = H + H G _3 H $ and
$    H _1 = H + H G _2 H $. 
The effective Hamiltonian in the $1$-space therefore
has the structure of a continued fraction:
\begin {equation} 
     \langle 1 \vert H_1 \vert 1 \rangle 
=  H_{11} + H_{12}  
    {\displaystyle 1 \over \displaystyle
     \omega - H_{22}      
   - H_{23}  
    {\displaystyle 1 \over \displaystyle
     \omega -  H_{33}  
   - H_{34} 
    {\displaystyle 1 \over \displaystyle
     \omega - H_{44}     
    } H_{43}  
    } H_{32} 
    } H_{21} 
 \ . \label {eq:514c} \end {equation} 
One can test the method of iterated resolvents in an almost 
trivial way: Assume that each of the blocks in 
Eq.(\ref{eq:508c})   has dimension 1. 
The matrix is thus a usual $4\times 4$-matrix. 
The inverse of a $1 \times 1$ `matrix' is trivial. 
The `effective Hamiltonian' is also a $1\times1$ `matrix' 
identical with the energy function $E(\omega)$.
For the matrix $\widetilde H _B$ in  Eq.(\ref{eq:508c})
one gets  simply
\begin {equation}   
    E(\omega) =  
    {\displaystyle 1\cdot 1  \over \displaystyle
     \omega - 2 - 
    {\displaystyle 3 \cdot 3 \over \displaystyle
     \omega - 4 - 
    {\displaystyle 5\cdot 5 \over \displaystyle
     \omega - 6 } } }
 \quad . \label {eq:516c} \end {equation}
As displayed in  Figure~\ref{fig:4_1}, the solutions of the
fixpoint  equation $ E(\omega) =  \omega  $  agree with the four 
eigenvalues to within computer accuracy. 
Note that a form like Eq.(\ref{eq:514c} ) or (\ref{eq:516c})
could  possibly be useful  to diagonalize a tridiagonal matrix of 
arbitraly large but finite dimension. 

\section{The Chains of the Effective Interaction} 
\label {sec:d}
The example in Appendix~\ref{sec:b} suggests
that the effective interaction wants to develop finite chains
of bare Hamiltonian blocks $\langle i\vert H \vert j\rangle$
alternating with resolvents of sector  Hamiltonians $H_n$.
This can be cast into a systematic procedure, as follows.
Sequential application of the recursion relation Eq.(\ref{eq:414}) 
allows to derive the useful relation 
\begin {equation}  
   H _n =  H + \sum_{m=n+1} ^N  H _m G _m  H _m
  \ . \label {eq:422} \end {equation}
To prove this, one writes down 
$     H _1 =  H _2 +  H _2 G _2  H _2 $ and 
$     H _2 =  H _3 +  H _3 G _3  H _3 $ and combines them 
to get  
$     H _1 =  H _3 +  H _2 G _2  H _2 
                  +  H _3 G _3  H _3 $. 
This is Eq.(\ref{eq:422}) for N=3. The general case is proven by
induction.
Eq.(\ref{eq:422}) relates the effective interaction in 
sector $n$ to the bare Hamiltonian $H$ and virtual 
scatterings into the higher  sectors  $ m > n $.  
This only {\em upward scattering} has important
consequences for the structure of the chains.  
How can one classify them? The  {\em number of propagators
in a chain} turns out to be a more useful criterium for that 
than, for example, the order of the coupling constant.
If a  chain has 3 propagators, it will contribute to  $ H ^{(3)} $. 
The bare interaction has no propagator and thus 
is the `chain' $ H ^{(0)} = H $.
The effective interaction is the sum of all possible chains:
\begin {equation}
    H _n = H ^{(0)} _n + H ^{(1)} _n + H ^{(2)} _n + H ^{(3)} _n 
         + H ^{(4)} _n + H ^{(5)} _n + \cdots
\ . \label {eq:426} \end {equation} 
The expansion is finite for any finite dimensional block matrix.
After its insertion into Eq.(\ref{eq:422}) one reads off 
{\em  upward recursion relations} with the general term
\begin {equation}
       H ^{(k+1)} _n = \sum \limits _{l>n} \ \Big( 
       H ^{(0)} _l G _l H ^{(k)} _l
     + H ^{(1)} _l G _l H ^{(k-1)} _l
     + \cdots 
     + H ^{(k-1)} _l G _l H ^{(1)} _l
     + H ^{(k)}   _l G _l H ^{(0)} _l
\ \Big) \ , \label {eq:436} \end {equation}
or explicitly for the first four of them
\begin {eqnarray}
       H ^{(1)} _n & = & \sum \limits _{l>n} \ \phantom{\Big(}
       H ^{(0)} _l G _l H ^{(0)} _l
\ \phantom{\Big)} \ , \label {eq:428} \\
       H ^{(2)} _n & = & \sum \limits _{l>n} \ \Big( 
       H ^{(0)} _l G _l H ^{(1)} _l
     + H ^{(1)} _l G _l H ^{(0)} _l
\ \Big) \ , \label {eq:430} \\
       H ^{(3)} _n & = & \sum \limits _{l>n} \ \Big( 
       H ^{(0)} _l G _l H ^{(2)} _l
     + H ^{(1)} _l G _l H ^{(1)} _l
     + H ^{(2)} _l G _l H ^{(0)} _l
\ \Big) \ , \label {eq:432} \\
       H ^{(4)} _n & = & \sum \limits _{l>n} \ \Big( 
       H ^{(0)} _l G _l H ^{(3)} _l
     + H ^{(1)} _l G _l H ^{(2)} _l
     + H ^{(2)} _l G _l H ^{(1)} _l
     + H ^{(3)} _l G _l H ^{(0)} _l
\ \Big) 
\ . \label {eq:434} \end {eqnarray}
\begin {table}[t]
\begin {center}
\caption [masses] {\label{tab:6_2} \em
   All chains of the effective interaction in the $q\bar q$-space 
   $H_1$  with one,  two and three propagators $G_i$ which are
   associated with the QCD matrix in 
   Figure~\protect\ref{fig:holy-1} 
   are enumerated.  The square   brackets for 
   $ H ^{(3)}$ refer to the three different sums in the defining 
   Eq.(\protect\ref{eq:442}). 
   } \vspace{1em}
\begin {tabular}  {||r|c||r|c||r|c||r|c||}
  \hline \hline    
    $\#$ & $ H _1^{(1)} $  and  $ H _1 ^{(2)} $    &
    $\#$ & $ H _1^{(3)}  [l_3,l_2,l_1] $  &  
    $\#$ & $ H _1 ^{(3)} [l_3,l_1,l_2] $  & 
    $\#$ & $ H _1 ^{(3)} [l_3,n_1,r_2] $ 
  \\ \hline \hline   
1 & $S G_2 S $   &
10&$F G_4 V G_3 V G_2 S $&15&$V G_3 S G_5 V G_2 S $&25&$V G_3 V G_2 V G_3 V$   
   \\
  2 & $V G_3 V $   & 
11&$F G_6 V G_3 V G_2 S $&16&$V G_3 V G_6 F G_2 S $&26&$F G_4 V G_3 V G_4 F$ 
   \\ 
  3 & $F G_4 F $   &  
12&$F G_6 S G_4 V G_3 V $&17&$F G_4 S G_6 V G_3 V $&27&$F G_6 F G_2 F G_6 F$
  \\ 
  4 & $F G_6 F $   &
13&$F G_6 V G_5 V G_2 S $&18&$F G_4 S G_6 F G_2 S $&28&$F G_6 V G_3 V G_6 F$ 
  \\ 
 5 & $V G_3 V G_2 S $ &
14&$F G_6 V G_5 S G_3 V $&19&$F G_4 V G_7 F G_3 V $&29&$F G_6 S G_4 S G_6 F$ 
  \\ 
 6 & $F G_4 V G_3 V $ &
    &                  &20&$F G_6 V G_7 V G_4 F $&30&$F G_6 V G_5 V G_6 F$ 
  \\ 
7 & $F G_6 F G_2 S $  &
   &                  &21&$F G_6 V G_7 F G_3 V $&31&$F G_6 F G_2 V G_3 V$ 
  \\ 
 8 & $F G_6 V G_3 V $ &
   &                  &22&$F G_6 S G_9 F G_2 S $&32&$F G_6 V G_3 V G_4 F$ 
  \\   
 9 & $F G_6 S G_4 F $ &
   &                  &23&$F G_6 V G_{10} F G_3 V $&     &   
  \\ 
    &                 &
   &                  &24&$F G_6 F G_{11} F G_4 F $&     & 
  \\ 
  \hline \hline   
\end {tabular}
\end {center}
\end {table}
For convenience, the chains up to order 4 are tabulated 
explicitly: 
\begin {eqnarray}
       H ^{(1)} _n & = & \sum \limits _{l_1>n} \ \phantom{\Big(} 
       H G _{l_1} H 
\ \phantom{\Big)} \ , \label {eq:438} \\
       H ^{(2)} _n & = & \sum \limits _{l_2>l_1>n} \ \Big(
       H G _{l_2} H G _{l_1} H + H G _{l_1} H G _{l_2} H 
\ \Big) \ , \label {eq:440} \end {eqnarray} 
\begin {equation} \begin {array} {lcrl}
       H ^{(3)} _n 
&=& \sum \limits _{l_3>l_2>l_1>n} \ \Big( & 
       H G _{l_3} H G _{l_2} H G _{l_1} H +
       H G _{l_1} H G _{l_2} H G _{l_3} H +
\\ & & &
       H G _{l_3} H G _{l_1} H G _{l_2} H +
       H G _{l_2} H G _{l_1} H G _{l_3} H 
\quad \Big) \\ 
&+& \sum \limits _{l_3>n_1>n \atop r_2>n_1>n } \ \Big( &
       H G _{l_3} H G _{n_1} H G _{r_2} H \quad \Big)
\ , \end {array} \label {eq:442} \end {equation}
\begin {equation} 
\begin {array} {lcrl}
       H ^{(4)} _n &=& \sum \limits _{l_4>l_3>l_2>l_1>n} \ \Big( &
       H G _{l_4} H G _{l_3} H G _{l_2} H G _{l_1} H +
       H G _{l_1} H G _{l_2} H G _{l_3} H G _{l_4} H +
\\  & & &
       H G _{l_4} H G _{l_1} H G _{l_2} H G _{l_3} H +
       H G _{l_2} H G _{l_1} H G _{l_3} H G _{l_4} H +
\\  & & &
       H G _{l_4} H G _{l_3} H G _{l_1} H G _{l_2} H +
       H G _{l_3} H G _{l_2} H G _{l_1} H G _{l_4} H +
\\  & & &
       H G _{l_4} H G _{l_2} H G _{l_1} H G _{l_3} H +
       H G _{l_3} H G _{l_1} H G _{l_2} H G _{l_4} H 
\quad \Big) \\ 
&+& \sum \limits _{l_4>l_3>l_2>n \atop l_4>l_3>r_1>n } \ \Big( &
       H G _{l_4} H G _{l_2} H G _{l_3} H G _{r_1} H +
       H G _{r_1} H G _{l_3} H G _{l_2} H G _{l_4} H 
\quad \Big) \\ 
&+& \sum \limits _{l_4>l_3>n \atop l_4>r_2>r_1>n } \ \Big( &
       H G _{l_3} H G _{l_4} H G _{r_1} H G _{r_2} H +
       H G _{r_2} H G _{r_1} H G _{l_4} H G _{l_3} H +
\\  & & &
       H G _{l_3} H G _{l_4} H G _{r_2} H G _{r_1} H +
       H G _{r_1} H G _{r_2} H G _{l_4} H G _{l_3} H 
\quad \Big) \ . \end {array} \label {eq:444} \end {equation}
\begin{figure} 
\vspace{-0.5mm}
\begin{minipage}[t]{65mm} \makebox[0mm]{}
\epsfxsize=65mm\epsfbox{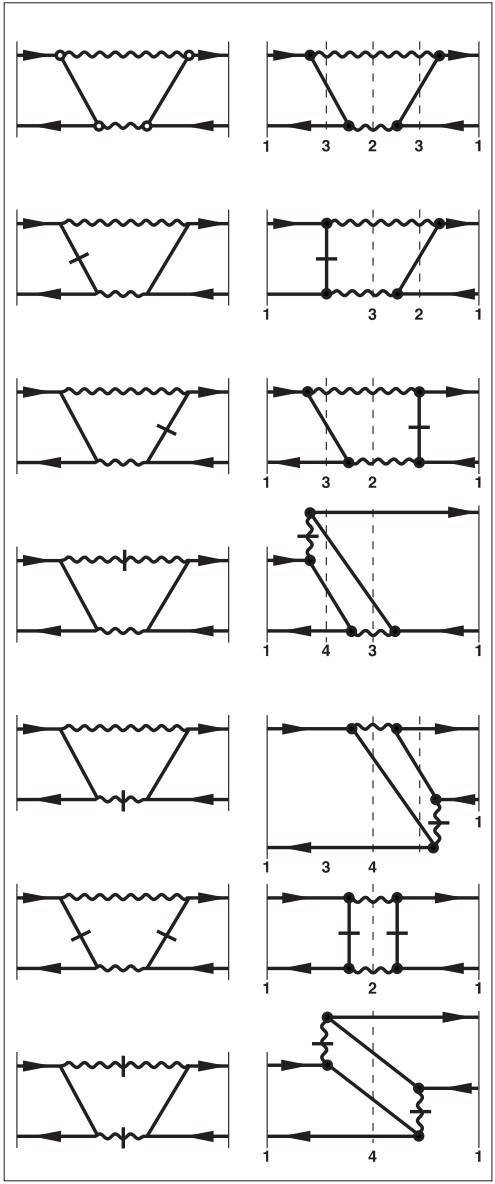}
\end{minipage} \hfill
\begin{minipage}[t]{95mm} \makebox[0mm]{}
\caption{\label{fig:6_6} \sl
The annihilation interaction with all instantaneous lines
explicitly inserted.
The annihilation graph 
of the effective interaction is 
displayed in the upper left of the figure.
It represents the following three-step procedure:  (1) A line
in between two interaction points is called a `physical' line 
representing the exchange of a physical particle; 
(2) Every physical line is the sum of a `dynamical' and an 
` instantaneous' line;  (3) Two instantaneous lines cannot
be connected at a vertex.  
Drawing  all possible diagrams with 
$0,1,2,\dots\ $ instantaneous lines like 
in  the left of the figure in the convention of 
Tang \protect\cite{tab91}, one obtains seven gaphs. 
On the right, 
the same  seven diagrams are drawn 
in the convention of Brodsky\protect\cite{brp91}. 
One can read off there the chains of different order, 
particularly   $ SG_2 S$, 
$ FG_4 F$, 
$ VG_3 VG_2 S$,
$ VG_4 VG_3 F$, and 
$ VG_3 VG_2 VG_3 V$.
When analyzed in the same way, the gluon exchange 
graph $U_{1,1}$ in Figure~\protect\ref{fig:6_1} provides 
one with the chain of order zero ($S$) and with $VG_3V$. 
All of them appear also in Table~\protect\ref{tab:6_2}.
But one misses a chain of order 1 which appears in the
Table, {\it i.e.} $ FG_6 F$.  Where is it, and where are the
other chains of higher order?~--- See the discussion in the
text.
}  \vfill \end{minipage}
\end{figure}
Table \ref{tab:6_2} enumerates  all chains up to 
order 3 for the full QCD-block-matrix as displayed in 
Figure~\ref{fig:holy-1}. For convenience, 
the type of interaction in the block matrix is accounted
for  by  $V$, $F$,  and $S$, refering to vertex-, fork-, and 
seagull-interactions, respectively. 

\section{The Gauge Trick}
\label{sec:f}

The present work has  been started keeping track explicitly
of the gauge remnants, the seagull  and the fork interactions
$S$ and $F$. 
With the work progressing it was realized that the formal
manipulations can be substantially simplified when one 
includes only the vertex interactions combined with the gauge
trick. Rather than formally, the argument  is lead here
diagrammatically, by way of example in 
Figure~\ref{fig:6_6} and its caption. 
Very instructively, one misses there a chain  of order 1 
which shows up in Table~\ref{tab:6_2}, particularly $ FG_6 F$.
It is found when analyzing the series in Eq.(\ref{eq:648}) 
term by term, some of which being displayed diagrammatically 
in Figure~\protect\ref{fig:6_4}. Consider the graph labelled 
(36b) in the Figure, standing for the chain
$V\overline G _3 V\,G_6\,V\overline G_3V$. 
Now, interprete the two outer gluon lines as instantaneous 
and write down the corresponding chain: it is the missing
$ FG_6 F$! The literally hundreds of graphs
analyzed diagrammatically  in preparing this work, 
cannot be discussed  here in detail, of course. 
Suffice it to state that {\em all chains with instantaneous 
interactions  as enumerated in Table~\ref{tab:6_2} 
have been identified explicitly as the instantaneous 
partners of dynamical chains like} 
$VGVGV\dots GVGVGV$. This should be sufficient 
evidence for the `gauge trick ' applied in this work.

\section{The cancellation of the gauge remnants} 
\label{sec:g}

Substituting the $\eta$-dependent terms of the 
spinor sum, Eq.(\ref{eq:7.2.08}), into the effective 
interaction yields straightforwardly
\begin {eqnarray} 
       \sum_{\lambda_g}
       \langle q ,\bar q \vert \,VR^\dagger\,
       \vert q^\prime, \bar q , g  \rangle 
       \langle q^\prime, \bar q ,g  \vert \,RV\,
       \vert q ^\prime ,\bar q ^\prime \rangle _\eta
       &=& - {n_c^2-1\over 2n_c}
        \,{g^2 r^2(Q,\Lambda) \over \Omega\,P^+}
        \,{(P^+)^2\over x_g k_g^+ }\times
\nonumber\\
        \times\bigg\{
        {\left[\overline u (q)\gamma_\mu k_g^\mu
        u(q^\prime)\right] \over  P^+\sqrt{x_{q}x_{q}^\prime}}
        {\left[\overline u (\bar q) \gamma_\nu\eta^\nu 
        u(\bar q ^\prime)\right]
        \over  P^+\sqrt{x_{\bar q} x_{\bar q}^\prime} }
        &+&{\left[ \overline u (q)\gamma_\mu\eta^\mu 
        u(q ^\prime)\right] \over P^+\sqrt{x_{q}x_{q}^\prime}}
        {\left[ \overline u (\bar q) 
        \gamma_\nu k_g^\nu u(\bar q ^\prime)\right]
        \over  P^+\sqrt{x_{\bar q} x_{\bar q}^\prime} } \bigg\} 
\  .   \end {eqnarray} 
The well-known property of the Dirac spinors
\begin{equation} 
        (k_q - k_q^\prime)^\mu\left[ 
        \overline u (k_q,\lambda_q) \,\gamma_\mu\,
        u(k_q^\prime,\lambda_q^\prime)\right] = 0
\      \end{equation} 
can be used for constructing the time-like null vectors
derived in section~\ref{sec:6}, {\it i.e.}
\begin{equation}      
       l_{\bar q} ^\mu = \left(k_g + k_{\bar q}^\prime 
       - k_{\bar q}\right) ^\mu
       = {\vec k _{g_{\!\perp}} ^{\,2}\over  2k_g^+}\ \eta ^\mu
       \quad{\rm and}\quad
       l^\mu_q = \left(k_g + k_q - k_q^\prime\right) ^\mu
       = {\vec k _{g_{\!\perp}} ^{\,2}\over  2k_g^+}\ \eta ^\mu
\ .   \end{equation}
Together with the resolvent
$\overline G _3 =-x_g/\vec k_{g_{\!\perp}}^{\,2}$
one gets most directly
\begin {eqnarray} 
       & & \sum_{\lambda_g} 
       \langle q ,\bar q \vert \,VR^\dagger\,
       \vert q^\prime, \bar q , g  \rangle 
       \overline G _3 
       \langle q^\prime, \bar q ,g  \vert \,RV\,
       \vert q ^\prime ,\bar q ^\prime \rangle _\eta =
\nonumber\\
       & & =  {n_c^2-1\over 2n_c}
        \,{g^2 r^2(Q,\Lambda) \over \Omega\,P^+}
        \,{1\over x_{ g}  ^2 }
        \ {\left[ \overline u (q) \gamma^+ u(q ^\prime)\right] 
        \over P^+\sqrt{x_q x_q^\prime} }
        {\left[ \overline u (\bar q) 
        \gamma^+ u(\bar q ^\prime)\right] 
        \over  P^+\sqrt{x_{\bar q} x_{\bar q}^\prime} } 
\  .   \end {eqnarray} 
Compare this with the seagull interaction evaluated 
directly by means of the tables \cite{brp91}
\begin {eqnarray} 
       \langle q ,\bar q \vert \,R^\dagger SR\,
       \vert q^\prime, \bar q^\prime \rangle 
       = - {n_c^2-1\over 2n_c}
       \ {g^2 r^2(Q,\Lambda)\over\Omega\,P^+}         
       \ {\left[ \overline u (k_q,\lambda_q) 
       \gamma^+ u(k_q^\prime,\lambda_q^\prime)\right]
       \over  \big( x_{q} - x_{q}^\prime\big) ^2
       \ P^+\sqrt{x_{q}x_{q}^\prime}}
       {\left[ \overline u (k_{\bar q},\lambda_{\bar q}) 
       \gamma^+ u(k_{\bar q}^\prime,\lambda_{\bar q}^\prime)
       \right]\over P^+ \sqrt{x_{\bar q} x_{\bar q}^\prime} }
\     \end {eqnarray} 
to conclude that {\em all gauge artefacts cancel each other
precisely}.
\end{appendix}

\end{document}